\documentclass[letterpaper,11pt]{article}

\usepackage{amsmath, amsthm, amssymb}
\usepackage[usenames, dvipsnames]{color}
\usepackage[normalem]{ulem} 
\usepackage{fullpage}
\usepackage{cite}
\usepackage[authoryear,round,longnamesfirst]{natbib}
\usepackage[noend]{algorithmic}

\usepackage{tikz, pgfplots}
\usepackage{enumerate}
\usepackage{hyperref}
\usepackage{xspace,color}
\usepackage{graphicx}
\usepackage{caption}
\usepackage{subcaption}
\usepackage{cleveref}
\usepackage{enumitem,linegoal}
\usepackage[linesnumbered,ruled,vlined]{algorithm2e}
\usepackage{comment}	
\usepackage{ctable}

\usepackage{mathtools}
\usepackage[flushleft]{threeparttable}
\usepackage{verbatim}
\usepackage{setspace}
\usepackage{bbm}
\usepackage{thm-restate}
\usepackage{footnote}
\usepackage{tcolorbox}
\usepackage{fdsymbol}
\usepackage{nicefrac}
\usepackage{array}
\usepackage{booktabs}
\usepackage{multirow}
\usepackage{listings}
\usepackage{apxproof}
\usepackage[framemethod=TikZ]{mdframed}
\newmdenv[
backgroundcolor=gray!5,
linecolor=black,
linewidth=0.7pt,
roundcorner=6pt,
innertopmargin=12pt,
innerbottommargin=12pt,
innerleftmargin=10pt,
innerrightmargin=10pt,
skipabove=12pt,
skipbelow=12pt
]{fancybox}

\makesavenoteenv{tabular}
\makesavenoteenv{table}
\crefname{enumi}{Condition}{Conditions}

\usepackage[margin=1in]{geometry}
\definecolor{darkgreen}{rgb}{0.0, 0.39, 0.0}

\hypersetup{
	colorlinks   = true, 
	urlcolor     = blue, 
	linkcolor    = blue, 
	citecolor   = darkgreen 
}

\definecolor{crimsonglory}{rgb}{0,0,0}

\newtheorem{example}{Example}

\newtheorem{theorem}{Theorem}[section]

\newtheorem{lemma}[theorem]{Lemma}

\newtheorem{definition}[theorem]{Definition}

\makeatletter
\def\GrabProofArgument[#1]{ #1: \egroup\ignorespaces}
\def\proof{\noindent\textbf\bgroup Proof%
	\@ifnextchar[{\GrabProofArgument}{. \egroup\ignorespaces}}

\makeatother

\usetikzlibrary{shapes.geometric, arrows.meta, positioning,shadows,calc,decorations.markings,patterns,graphs,mindmap, decorations.pathmorphing,fit,shapes.misc,backgrounds}

\tikzstyle{startstop} = [rectangle, rounded corners, minimum width=3cm, minimum height=1cm, text centered, draw=black, fill=red!30]
\tikzstyle{process} = [rectangle, minimum width=3cm, minimum height=1cm, text centered, draw=black, fill=orange!30]
\tikzstyle{decision} = [diamond, minimum width=3cm, minimum height=1cm, text centered, draw=black, fill=green!30]
\tikzstyle{arrow} = [thick,->,>=stealth]
\tikzset{
agentj/.style={circle, draw, minimum size=7mm, font=\small, inner sep=1pt, fill=agentj}, 
agenti/.style={circle, draw, minimum size=7mm, font=\small, inner sep=1pt, fill=agenti}, 
bundle/.style={rectangle, draw=orange!70!black, rounded corners,
minimum width=0.7cm, minimum height=0.5cm, fill=orange!30, opacity=1, font=\scriptsize, align=center},
heavy/.style={-{Stealth[length=2mm]}, thick, red, bend left=15},
heavyrev/.style={-{Stealth[length=2mm]}, thick, red, bend right=15}, 
light/.style={-{Stealth[length=2mm]}, thick, green!70!black, bend left=15},
rotation/.style={-{Stealth[length=2mm]}, thick, magenta},
		choice/.style={->, dashed},
subpool/.style={rectangle, draw=blue!70!black, thick,
	rounded corners, minimum width=1cm, minimum height=0.6cm,
	fill=white, opacity=1, font=\scriptsize, align=center},
smallset/.style={rectangle, draw=green!70!black, thick,
	rounded corners, minimum width=0.8cm, minimum height=0.4cm,
	fill=green!20, opacity=1},
poolimg/.style={inner sep=0pt, anchor=center, opacity=0.8, minimum width=2.5cm, minimum height=1.0cm},
rotind/.style={circle, draw=white, thick, minimum size=8mm},
ret/.style={    decorate,
	decoration={
		snake,               
		amplitude=1.5pt,     
		segment length=6pt,  
		post length=7pt,     
		pre length=3pt       
	},
	thick,                 
	-{Stealth[length=2mm]},
	blue  },
	good/.style={
		rectangle,
		draw=black!80,
		fill=white,
		rounded corners=3pt,
		font=\small\bfseries,
		inner sep=2pt,
		text=black
	}
}
\definecolor{skyblue}{RGB}{135,206,235}  
\definecolor{coral}{RGB}{255,127,80}   
\definecolor{limegreen}{RGB}{50,205,50}    
		\definecolor{agent1}{RGB}{173,216,230}
\definecolor{agent2}{RGB}{255,182,193}
\definecolor{agent3}{RGB}{152,251,152}

\usepackage[latin1]{inputenc}

\newcommand{\efx}{\mathsf{EFX}}

\newcommand{\valu}{v}

\newcommand{\invsq}{\tfrac{1}{\sqrt{2}}}
\newcommand{\goods}{M}

\newcommand{\agents}{N}

\newcommand{\sagents}{\agents^*}
\newcommand{\ragents}{\tilde\agents}
\newcommand{\alloc}{X}
\newcommand{\ralloc}{\tilde X}
\newcommand{\salloc}{X^*}
\DeclareRobustCommand{\rchi}{{\mathpalette\irchi\relax}}
\newcommand{\irchi}[2]{\raisebox{\depth}{$#1\chi$}} 
\newcommand{\malloc}{\rchi}

\newcommand{\pool}{\mathcal{P}}

\newcommand{\good}{g}

\newcommand{\relevant}[2]{#1{\scriptstyle{\langle#2\rangle}}}

\newcounter{proccnt}

\newcommand{\konote}[1]{}

\usepackage[normalem]{ulem} 

\title{Improved Approximate EFX Guarantees for Multigraphs}

\author{
	Alireza Kaviani\thanks{Sharif University of Technology, Tehran, Iran}
	\and
	Alireza Keshavarz \footnotemark[1] \and
	Masoud Seddighin\thanks{Tehran Institute for Advanced Studies (TeIAS), Tehran, Iran}
	\and
	AmirMohammad Shahrezaei\footnotemark[1]
}

			\definecolor{darkgreen}{rgb}{0.0, 0.5, 0.0}
		\definecolor{agentk}{RGB}{152,251,152}  
\definecolor{agenti}{RGB}{173,216,230}  
\definecolor{agentj}{RGB}{255,182,193}  
\begin{document}
	\newcommand{\ignore}[1]{}
	\renewcommand{\theenumi}{\roman{enumi}.}
	\renewcommand{\labelenumi}{\theenumi}
	\sloppy
	\date{} 
	\newenvironment{subproof}[1][\proofname]{
		\renewcommand{\Box}{ \blacksquare}%
		\begin{proof}[#1]%
		}{%
		\end{proof}%
	}
	\maketitle
	
	\thispagestyle{empty}
	\allowdisplaybreaks

	\vspace{0.5cm}
	\begin{abstract}
		In recent years, a new line of work in fair allocation has focused on $\efx$ allocations for $(p, q)$-bounded valuations, where each good is relevant to at most $p$ agents, and any pair of agents share at most $q$ relevant goods. For the case $p = 2$ and $q = \infty$, such instances can be equivalently represented as multigraphs whose vertices are the agents and whose edges represent goods, each edge incident to exactly the one or two agents for whom the good is relevant. A recent result of \citet{amanatidis2024pushing} shows that for additive $(2,\infty)$ bounded valuations, a $(\nicefrac{2}{3})$-$\efx$ allocation always exists. In this paper, we improve this bound by proving the existence of a $(\nicefrac{1}{\sqrt{2}})$-$\efx$ allocation for additive $(2,\infty)$-bounded valuations.

	\end{abstract}
	\section{Introduction}\label{introduction}

Fair division is a fundamental problem with deep connections to mathematics, economics, and  computer science. The central objective is to allocate a resource among a set of agents so that each perceives their share as fair, according to some well-defined notion of fairness. A canonical example is the cake-cutting problem, where the resource is modeled as a divisible, heterogeneous good.

The history of cake cutting includes many interesting and insightful results. One of the most classic result is the existence of envy-free divisions, where no agent prefers another's share over their own. Such divisions are guaranteed to exist under mild assumptions on the valuation functions.

A key feature that makes the cake-cutting problem tractable is the divisibility of the resource---it can be cut at any point. However, when moving from divisible resources to indivisible goods---items that cannot be split---the complexity of fair allocation increases significantly. In this setting, the goal is to distribute a set of goods $\goods$ among $n$ agents, where each agent $i$ has a valuation function $\valu_i: 2^\goods \to \mathbb{R}^+$ that assigns a non-negative value to each subset of goods. Unlike in the divisible case, envy-free allocations are not guaranteed to exist; for example, consider a single valuable good and two agents.

For indivisible goods, several relaxations of envy-freeness have been proposed to make the concept more applicable. These relaxations typically allow an agent to envy another only if the envy can be eliminated by removing a single item from the envied bundle. The removed item may be the most valuable one ($\textsf{EF1}$), the least valuable one ($\efx$), a randomly chosen item (\textsf{EFR}), one less preferred item ($\textsf{EFL}$), or defined in some other way. Among these, $\efx$ is widely regarded as the strongest and most compelling relaxation.

\textit{Envy-Freeness up to Any Good} (\(\efx\)) was introduced by \citet{caragiannis2016unreasonable}. Formally, consider an allocation \(\alloc = (\alloc_1, \alloc_2, \ldots, \alloc_n)\), where \(\alloc_j\) denotes the bundle of goods allocated to agent \(j\). The allocation \(\alloc\) is \(\efx\) if, for every pair of agents \(i\) and \(j\), and for every good \(\good \in \alloc_j\), it holds that
$
\valu_i(\alloc_i) \geq \valu_i(\alloc_j \setminus \{\good\}).
$
In other words, after removing any single good from another agent's bundle, agent \(i\) values their own allocation at least as much as the adjusted bundle.

Perhaps $\efx$ is the strongest fairness guarantee one can hope for when allocating indivisible goods. 
The existence of \(\efx\) allocations remains a major open problem, and it is still unclear whether \(\efx\) allocations exist for every instance involving indivisible goods. However, researchers have made progress in specific cases. For example, \(\efx\) allocations are known to exist for two agents with monotone valuations \citep{plaut2020almost} and for three agents, where two have monotone valuations and one has an \textsf{MMS}-feasible valuation \citep{akrami2023efx}. Moreover, in certain settings, relaxations of \(\efx\) can be guaranteed, such as a \(0.618\)-approximate \(\efx\) allocation for additive valuations \citep{amanatidis2020multiple} and an \(\efx\) allocation that allows discarding up to \(n-1\) goods under monotone valuations \citep{chaudhury2021little}. Nonetheless, finding a general method for complete \(\efx\) allocations remains open.

In this paper, we push the boundaries of the results for $\efx$ by proving the existence of a \((\nicefrac{1}{\sqrt{2}})\)-\(\efx\) allocation for additive \((2, \infty)\)-bounded instances. The concept of \((p, q)\)-bounded valuations, recently introduced by \citet{christodoulou2023fair}, refers to scenarios where each good has a nonzero value (is relevant) for at most \(p\) agents, and any pair of agents shares at most \(q\) common relevant goods. \citet{christodoulou2023fair} first demonstrated that \((2, 1)\)-bounded instances always admit an \(\efx\) allocation. \citet{amanatidis2024pushing} later extended this result to \((2, \infty)\)-bounded instances, achieving a \((\nicefrac{2}{3})\)-\(\efx\) allocation. \citet{kaviani2024almost} subsequently showed that complete \(\efx\) allocations exist for \((2, \infty)\)-bounded instances when the valuation functions are restricted additive --- where each good $g$ is valued at either $0$ or \(v_g\) by each agent. Additionally, \citet{afshinmehr2024efx} provided complete \(\efx\) allocations for the case where the underlying relevance graph is a bipartite multigraph.
Here, we extend these results by establishing a \((\nicefrac{1}{\sqrt{2}})\)-\(\efx\) allocation for $(2,\infty)$-bounded valuations.

	\subsection{Further Related Work}

For the cake-cutting problem, comprehensive overviews are provided in the surveys by \citet{brams1996cake} and \citet{procaccia2015cake}. In the context of indivisible goods, several fairness notions have been proposed, including the maximin share guarantee \citep{Procaccia:first}, \textsf{EF1} \citep{budish2011combinatorial}, \(\efx\) \citep{budish2011combinatorial,caragiannis2016unreasonable}, \textsf{EFL} \citep{barman2018groupwise}, and \textsf{EFR} \citep{farhadi2021almost}, $\textsf{EEFX}$ \cite{akrami2024epistemic}.  For a broader discussion of fairness notions, we refer the reader to the surveys by \citet{amanatidis2022fair} and \citet{aziz2022algorithmic}.

Despite its theoretical appeal, the existence of \(\efx\) allocations remains an unresolved problem. However, notable progress has been made in specific cases \citep{chan2019maximin,babaioff2021fair,amanatidis2021maximum,plaut2020almost,chaudhury2024efx,akrami2023efx,akrami2022ef2x,ashuri2024ef2xexistsagents,ghosal2024efx}. 
Another active research area explores \(\efx\) allocations with charity, where some goods remain unallocated. This line of research was initiated by \citet{caragiannis2019envy}, who proposed an allocation that satisfies $\efx$ and guarantees a Nash welfare of at least half the maximum possible.
\citet{chaudhury2021little} showed that such allocations can exist with up to \(n-1\) unallocated goods, and \citet{berger2022almost} improved this by reducing the number to \(n-2\). 
\citet{akrami2022ef2x} further demonstrated that \(\efx\) allocations can be achieved with at most \(\lfloor \nicefrac{n}{2} \rfloor - 1\) unallocated items under restricted additive valuations. Recent work by \citet{kaviani2024almost} extended this to \((\infty, 1)\)-bounded valuations, with a similar number of unallocated items. Additionally, some studies have connected approximate \(\efx\) allocations with a sublinear number of discarded goods to a combinatorial problem called the rainbow cycle number \cite{chaudhury2021improving,akrami2023efx,jahan2023rainbow}.

	\section{Preliminaries}

A fair allocation instance involves three elements: agents, goods, and valuations. The agents are represented by \(\agents = \{1, 2, \dots, n\}\), and the goods by \(\goods\). Agent \(i\) has a function \(\valu_i: 2^\goods \rightarrow \mathbb{R}^{\geq 0}\) that assigns a non-negative value to every subset of goods. These functions are monotone, meaning that for any subsets \(S\) and \(T\) of \(\goods\), if \(S \subseteq T\), then \(\valu_i(S) \leq \valu_i(T)\). In addition, we assume that valuations are additive; that is, for every pair of disjoint subsets \( S \) and \( T \), we have \( \valu_i(S \cup T) = \valu_i(S) + \valu_i(T) \). 

We say that a good \( g \) is \emph{irrelevant} to agent \( i \) if \( \valu_i(\{g\}) = 0 \); otherwise, we call \( g \) \emph{relevant} to \( i \).
A family of valuation functions is \textit{\((p, q)\)-bounded} if each good is relevant to at most \( p \) agents, and each pair of agents shares at most \( q \) relevant goods. In this paper, we focus on the special case where \( p = 2 \) and \( q = 1 \). 
For any subset \( S \subseteq \goods \), let \( \relevant{S}{i} \) denote the set of goods in \( S \) that are relevant to agent \( i \), and let \( \relevant {S}{i, j} \) denote the set of goods in \( S \) that are relevant to both  \( i \) and \( j \). Note that \( \relevant{S}{i, i} \) refers to the set of goods that are relevant \emph{only} to agent \( i \), and is therefore different from \( \relevant{S}{i} \).

An \emph{allocation} $\alloc$ is a distribution of goods (not necessarily all) into $n$ bundles $( \alloc_1, \alloc_2, \dots, \alloc_n )$, where $\alloc_i$ denotes the bundle allocated to agent~$i$. In this paper, we use $\pool$ to represent the set of unallocated goods, referred to as the \emph{pool}. An allocation is \emph{complete} if $\pool = \emptyset$, and \emph{partial} otherwise.

Let $\beta \in [1, +\infty]$ be a constant. An agent $i$ $\beta$-envies  agent $j$ if $\beta \valu_i(\alloc_i) < \valu_i(\alloc_j)$. Furthermore, an agent $i$ $\beta$-strongly envies  agent $j$ if there exists a good $g \in \alloc_j$ such that $\beta \valu_i(\alloc_i) < \valu_i(\alloc_j \setminus \{g\})$. For $\alpha \in [0, 1]$, an allocation is $\alpha$-$\efx$ if no agent $(\nicefrac{1}{\alpha})$-strongly envies another.

Our algorithm relies on the concept of the \textit{weighted envy graph}, first introduced by \citet{farhadi2021almost}. For an allocation \( \alloc \), the weighted envy graph \( G_\alloc \) is a complete weighted directed graph where each vertex represents an agent. For any pair of agents \( i \) and \( j \), there is a directed edge from  \( i \) to  \( j \) with weight \( w_\alloc(i, j) = \nicefrac{\valu_i(\alloc_j)}{\valu_i(\alloc_i)} \). Since agents and vertices correspond one-to-one, we use the terms interchangeably and may refer to a vertex by its associated agent.
For any $\alpha \in [0, \infty]$, we define $G_{\alpha, \alloc}$ as the subgraph of $G_\alloc$ that includes only edges with a weight greater than $\alpha$. For example, $G_{1, \alloc}$ represents the envy graph where an edge from agent $i$ to agent $j$ indicates that agent $i$ envies agent $j$, while $G_{0, \alloc}$ includes an edge if agent $j$'s bundle has a non-zero value for agent $i$.

In this paper, we classify the edges of \(G_\alloc\) into two categories: an edge \((i,j)\) is called \emph{\textbf{heavy}} if \(w_\alloc(i,j) > \nicefrac{1}{\sqrt{2}}\), and \emph{\textbf{light}} otherwise.
 
\begin{figure}[t]
	\centering
	\scalebox{0.7}{
		\begin{tikzpicture}[
			node distance=1.5cm,
			agent/.style={
				circle,
				draw=black,
				minimum size=10mm,
				font=\large\bfseries,
				inner sep=1pt,
				fill=gray!20
			},
			good/.style={
				circle,
				draw=black,
				fill=yellow!20,
				rounded corners=3pt,
				minimum size=7mm,
				font=\scriptsize\bfseries\color{black}
			},
			wedge/.style={
				-{Stealth[length=3mm, width=3mm]},
				line width=1.2pt,
				color=black!70  
			},
			weight/.style={
				font=\footnotesize,
				inner sep=2pt,
				fill=white,
				rounded corners=2pt
			},
			good/.style={
				rectangle,
				draw=black!80,
				fill=white,
				rounded corners=3pt,
				font=\small\bfseries,
				inner sep=2pt,
				text=black
			}
			]
			
			\begin{scope}[xshift=-7cm]
				\node[agent, fill=agentj] (a1) at (0,0) {1};
				\node[agent, fill=agentj] (a2) at (4,0) {2};
				\node[agent, fill=agentj] (a3) at (2,-3) {3};
				
				\draw[dashed, color=gray] 
				(a1) .. controls (1,1) and (3,1) .. node[good,solid] {$g_1$} (a2)
				node[ color=black, pos=0.3] {2}  
				node[ color=black, pos=0.7] {8}; 
				
				\draw[ dashed, color=gray] 
				(a1) .. controls (1,-1) and (3,-1) .. node[good,solid] {$g_2$} (a2)
				node[ color=black, pos=0.3] {3}
				node[ color=black, pos=0.7] {3};
				
				\draw[ dashed,color=gray] 
				(a1) .. controls (0.5,-1.5) and (1.5,-2.5) .. node[good, pos=0.4,solid] {$g_3$} (a3)
				node[ color=black, pos=0.2] {1}
				node[ color=black, pos=0.7] {4};
				
				\draw[ dashed, color=gray] 
				(a2) .. controls (3.5,-1.5) and (2.5,-2.5) .. node[good, pos=0.4,solid] {$g_4$} (a3)
				node[ color=black, pos=0.2] {4}
				node[ color=black, pos=0.7] {3};
			\end{scope}
			
			\begin{scope}[yshift=0cm]
				\node[agent, fill=agentj] (a1) at (0,0) {1};
				\node[agent, fill=agentj] (a2) at (4,0) {2};
				\node[agent, fill=agentj] (a3) at (2,-3) {3};
				
				\draw[light] (a2) -- node[weight] {$\frac{3}{8}$} (a1);
				\draw[heavy] (a3) -- node[weight, pos=0.4] {$\frac{4}{3}$} (a1);
				\draw[light] (a2) -- node[weight, pos=0.4] {$\frac{4}{8}$} (a3);
				\draw[light] (a1) to[out=45,in=135] node[weight] {$\frac{2}{4}$} (a2);
			\end{scope}
			
			\begin{scope}[xshift=7cm,yshift=0cm]
				\node[agent, fill=agentj] (b1) at (0,0) {1};
				\node[agent, fill=agentj] (b2) at (4,0) {2};
				\node[agent, fill=agentj] (b3) at (2,-3) {3};
				
				\draw[heavy] (b3) -- node[weight] {$\frac{5}{3}$} (b1);
			\end{scope}
			
			\node[bundle] (x1) at (2,-5) {
				\begin{tabular}{@{}c@{}}
					\textbf{X\textsubscript{1}} \\
					\tikz\node[good]{$g_2$}; \tikz\node[good]{$g_3$};
				\end{tabular}
			};
			
			\node[bundle] (x2) at (6,-5) {
				\begin{tabular}{@{}c@{}}
					\textbf{X\textsubscript{2}} \\
					\tikz\node[good]{$g_1$};
				\end{tabular}
			};
			
			\node[bundle] (x3) at (10,-5) {
				\begin{tabular}{@{}c@{}}
					\textbf{X\textsubscript{3}} \\
					\tikz\node[good]{$g_4$};
				\end{tabular}
			};
			
			\node[above] at (2, 2) {\large $G_{0,X}$};
			\node[above] at (9,2) {\large $G_{1,X}$};
			
	\end{tikzpicture}}
	\caption{An example of an allocation and its weighted envy graph. See Table \ref{tab:styleguide} for style guidance.}
	\label{fig:graph}
\end{figure}
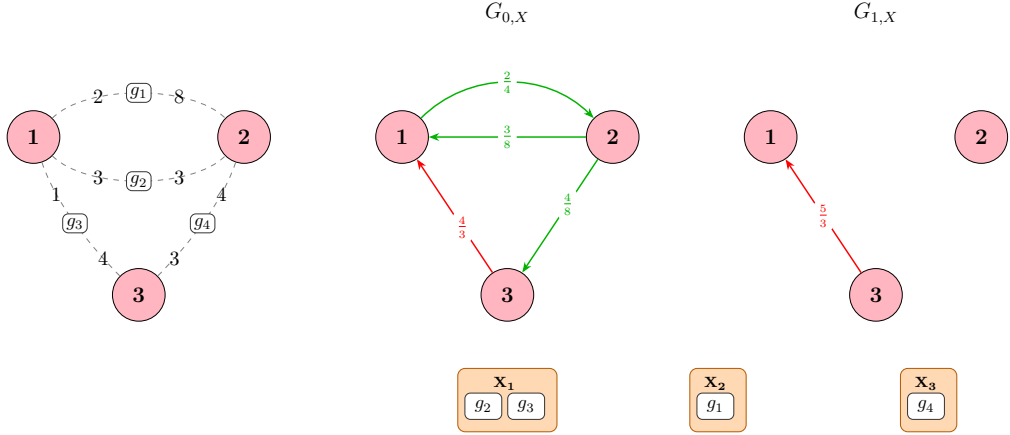

\begin{example}

Consider the instance shown on the left of \Cref{fig:graph}. Here, \(g_1\) is relevant to agents 1 and 2, \(g_3\) to 1 and 3, and \(g_4\) to 2 and 3. For agent 2,
$
v_2(\{g_1\}) = 8, v_2(\{g_2\}) = 3,\quad v_2(\{g_3\}) = 0, v_2(\{g_4\}) = 4.
$
The figure also shows an allocation \(X\) and the graphs \(G_{0,X}\) and \(G_{1,X}\). For example, since
$
v_3(X_3) = v_3(\{g_4\}) = 3, 
v_3(X_1) = v_3(\{g_2\}) + v_3(\{g_3\}) = 0 + 4 = 4,
$
we have \(w_X(3,1) = 4/3 > 1\), so this edge appears in \(G_{1,X}\). Thus agent 3 envies and strongly envies agent 1, but does not \(\sqrt{2}\)-strongly envy her. One can verify that this allocation is \((\nicefrac{1}{\sqrt{2}})\)-\(\efx\).
\end{example}
\paragraph{Structure of our Algorithm.}
Throughout the algorithm, we maintain the pairs $\langle \ragents, \ralloc \rangle$ and $\langle \sagents, \salloc \rangle$, along with the pool of unallocated goods $\pool$. Here, $\ragents$ and $\sagents$ partition the set of agents $\agents$, representing the remaining and finalized agents, respectively. Allocations $\ralloc$ and $\salloc$ allocate goods to agents in $\ragents$ and $\sagents$. Together, $\ralloc$, $\salloc$, and $\pool$ form a partition of the goods: each good is either allocated to exactly one agent in $\ragents$ or $\sagents$, or remains in $\pool$.
We begin with $\ragents = \agents$, $\sagents = \emptyset$, and $\salloc$ empty. The initial allocation $\ralloc$ assigns one good to each agent to maximize the Nash Social Welfare (i.e., the geometric mean of the agents' values) over all such allocations. This is called the \emph{basic feasible allocation}. By a result of \citet{kaviani2024almost}, we can assume that every agent receives a good of positive value in this allocation. Therefore, initially, $\ralloc$ assigns exactly one good to each agent in $\ragents$, and the remaining $m - n$ goods are in $\pool$.
The algorithm ends when $\ragents = \emptyset$ and $\sagents = \agents$. At that point, $\ralloc$ and $\pool$ are empty, and $\salloc$ gives the final allocation. 

\begin{toappendix}
\section{Style Guidline}
	\begin{table}[h!]
		\centering
		\caption{Figures Style Guide}
		\label{tab:styleguide}
		\begin{tabular}{>{\centering}m{2cm} >{\centering}m{3cm} m{6cm} m{3cm}}
			\toprule
			\textbf{Element} & \textbf{Visual Sample} & \textbf{Description}  \\
			\midrule
			
			\multirow{2}{*}{Agents} 
			& \begin{tikzpicture}[baseline=-3mm]
				\node[agentj] {$j$};
			\end{tikzpicture} 
			& Remaining agent in $\ragents$ & 
\\

			\cline{2-4}
			& \vspace{1mm}\begin{tikzpicture}[baseline=-3mm]
				\node[agenti] {$i$};
			\end{tikzpicture} 
			& Finalized agent in $\sagents$& 
\\
			\midrule
			\multirow{2}{*}{Bundles} 
			& \begin{tikzpicture}[baseline=-3mm]
				\node[bundle] {$\ralloc_i$};
			\end{tikzpicture} 
			& Allocated bundle & 
\\
			\cline{2-4}
			& \vspace{1mm} \begin{tikzpicture}[baseline=-3mm]
			\node[bundle,fill =green!20,draw=black, thick] {$\ralloc_i$};
			\end{tikzpicture} 
			& Finalized allocated bundle & 
\\
			
			\midrule
			\multirow{3}{*}{Edges} 
			& \vspace{1mm}\begin{tikzpicture}[baseline=-3mm]
				\draw[heavy] (0,0) -- (1,0);
			\end{tikzpicture} 
			& Heavy edge & 
 \\
			
			\cline{2-4}
			& \vspace{2mm} \begin{tikzpicture}[baseline=-3mm]
				\draw[light] (0,0) -- (1,0);
			\end{tikzpicture} 
			& Light edge & 
 \\
			
			\cline{2-4}
			& \vspace{2mm}\begin{tikzpicture}[baseline=-3mm]
				\draw[choice] (0,0) -- (1,0);
			\end{tikzpicture} 
			& Compare bundles & 
\\
			
			\bottomrule
		\end{tabular}
	\end{table}
\end{toappendix}
	\section{Overview of the Algorithm and Ideas}\label{sec:resultstech}

We prove the existence of a complete $(\nicefrac{1}{\sqrt{2}})$-$\efx$ allocation for cases where valuation functions are additive and $(2, \infty)$-bounded. 
Our algorithm consists of five updating rules followed by a final step. At each step, we apply the first applicable rule in sequence. Thus, if the $k$-th rule is applied at a given step, it implies that the first $k-1$ rules were not applicable at that step. Each rule modifies $\ralloc$ for some agents in $\ragents$, possibly finalizing their bundles and transferring them to $\sagents$. Once an agent is finalized, her allocation in $\salloc$ remains unchanged for the rest of the process. 
The rules are carefully designed to maintain five key properties throughout the algorithm:

\newcommand{\propertylabel}[2]{\item[\textbf{#1}] \label{prop:#2}}

\begin{fancybox}
	\begin{description}
		\propertylabel{(\textsf{i}) }{appxenvy}
		Allocation $\salloc$ is $(\nicefrac{1}{\sqrt 2})$-$\efx$ for the agents in $\sagents$. 
		\propertylabel{(\textsf{ii}) }{unionef}
For every agent $i \in \sagents$, she does not $\sqrt{2}$-envy the union of the bundles of the remaining agents and the pool. In other words, $\valu_i(\salloc_i) \geq \valu_i\left(\bigcup_{j \in \ragents } \ralloc_j \cup \pool\right) / \sqrt{2}.$

		\propertylabel{(\textsf{iii}) }{remainingefx}
		Allocation $\ralloc$  is $\efx$ for the agents in $\ragents$.
		\propertylabel{(\textsf{iv}) }{finalizedefx}
		For every agent $i \in \ragents$, she does not strongly envy any agent $j \in \sagents$. In other words, for every 
		$
		\good \in \salloc_j$ we have $ \valu_i(\ralloc_i) \geq \valu_i \left( \salloc_j\setminus \{g\}\right).
		$
		\propertylabel{(\textsf{v}) }{singletype}
		For each remaining agent $i \in \ragents$, there exists an agent $j$ such that $\malloc_i \subseteq \relevant{\goods}{i,j}$.
	\end{description}
\end{fancybox}

\begin{figure}[t]
	\centering
	\scalebox{0.7}{
	\begin{tikzpicture}[
		node distance=3cm,
		agent/.style={
			circle,
			draw=black,
			minimum size=10mm,
			font=\large\bfseries,
			inner sep=1pt
		},
		arrow/.style={
			-{Stealth[length=3mm]},
			line width=1pt,
			color=black
		},
		allocationbox/.style={  
			rounded rectangle,
			rounded corners=15pt,
			thick,
			dashed,
			inner sep=14mm,  
		}
		]
		

		\node[agent, fill=agenti] (i) at (2,0) {$\sagents$};
		\node[agent, fill=agentj] (j) at (10,0) {$\ragents$};
		
		\begin{scope}[on background layer]
			\node[allocationbox, 
			draw=green!120, 
			fill=green!20,  
			fit=(i)] (salloc) {};
			\node[allocationbox, 
			draw=orange!60, 
			fill=orange!20,  
			fit=(j)] (ralloc) {};
		\end{scope}
		
		\node[above=2mm of salloc, font=\large, orange!80!black] {$\salloc$};
		\node[above=2mm of ralloc, font=\large, blue!70!black] {$\ralloc \cup \pool$};
		
		\draw[arrow,color=red] (i) to[out=60,in=180] 
		node[pos=0.7, above] {\footnotesize\textsf{Prop.(ii)}} (ralloc);
		
		\draw[arrow] (i) to[out=135,in=225,looseness=8] 
		node[pos=0.5, left] {\footnotesize\textsf{Prop.(i)} } (i);
		
		\draw[arrow, dashed] (j) to[out=-45, in=45, looseness=8] 
		node[pos=0.5, right] {\footnotesize\textsf{Prop.(iii)}} (j);
		
		\draw[arrow, dashed] (j) to[out=-120,in=-60] 
		node[pos=0.5, above] {\footnotesize\textsf{Prop.(iv)}} (i);
	\end{tikzpicture}}
	\caption{A schematic overview of the properties.
		Dashed arcs: No Strongly envy. Solid black arc: no $\sqrt{2}$-strongly envy, solid red arc: no $\sqrt{2}$- envy.}
	\label{fig:kij_flow}
\end{figure}
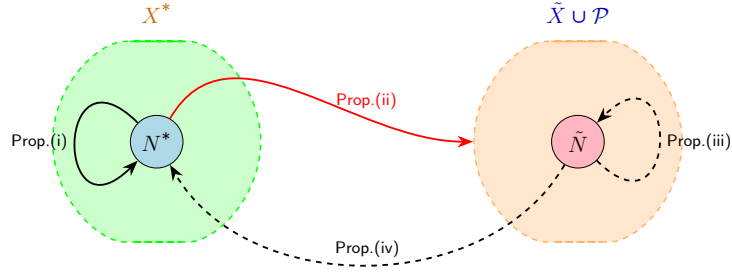

A graphical overview of the first four properties is shown in \Cref{fig:kij_flow}.  
Property~\textsf{(i)} ensures that the final allocation is $(\nicefrac{1}{\sqrt{2}})$-$\efx$.  
Property~\textsf{(ii)} allows us to update $\ralloc$ without concern for the agents in $\sagents$, as it guarantees that no agent in $\sagents$ will $\sqrt{2}$-envy the combined set of goods in $\ralloc$ and $\pool$.  
Properties~\textsf{(iii)} and~\textsf{(iv)} help establish properties~\textsf{(i)} and~\textsf{(ii)} when an agent is moved from $\ragents$ to $\sagents$.  
Property~\textsf{(v)} plays a slightly different role. While it is not directly used to prove the $(\nicefrac{1}{\sqrt{2}})$-$\efx$ guarantee, it is instrumental in maintaining the other properties during the update and finalization steps of the algorithm. It ensures that for every remaining agent~$i$, there exists an agent~$j$ (possibly~$i$ itself) such that all goods in~$\ralloc_i$ are relevant only to~$i$ and~$j$; that is, $\ralloc_i \subseteq \relevant{\goods}{i,j}$. This implies that the goods allocated to~$i$ matter to at most one other agent, which greatly simplifies the structure of~$G_{0,\ralloc}$: each vertex has at most one incoming edge.

These properties hold trivially for the basic feasible allocation. Below, we briefly describe the updating rules. Each rule is accompanied by a visual illustration. For better understanding these figures, see Table~\ref{tab:styleguide} in the appendix.

\paragraph{Rule 1.}  
The first updating rule removes envy toward goods of the same type in the pool. Specifically, if there exist agents \(i\) and \(j\) such that \(i\) envies \(\relevant{\pool}{i, j}\), the rule is applicable. In that case, we identify a minimal subset of \(\relevant{\pool}{i, j}\) that is still envied by either \(i\) or \(j\). Note that these goods have zero value to all other agents. We then allocate this subset to the agent who envies it (arbitrarily choosing one if both do), and return their current bundle to the pool. Once this rule is no longer applicable, it means that for every pair \(i, j\), the set \(\relevant{\pool}{i, j}\) is not envied by either agent. Thus, any remaining envy toward \(\relevant{\pool}{i}\) must come from agent \(i\) alone.
\vspace{-0.3cm}
\begin{figure}[h!]
	\centering
	\scalebox{0.8}{
	\begin{tikzpicture}[
		panelLabel/.style={font=\bfseries\footnotesize, anchor=west},
		poolimg/.style={inner sep=0pt, anchor=center, opacity=0.8},
		subpool/.style={rectangle, draw=blue!70!black, thick,
			rounded corners, minimum width=0.9cm, minimum height=0.7cm,
			fill=white!30, opacity=1},
		subset/.style={rectangle, draw=green!40!black, thick, dashed,
			rounded corners, minimum width=0.7cm, minimum height=0.45cm,
			fill=blue!60, opacity=1},
		agent/.style={circle, draw, minimum size=8mm, font=\small, inner sep=1pt},
		bundle/.style={rectangle, draw=orange!70!black, rounded corners,
			minimum width=0.65cm, minimum height=0.45cm,
			fill=orange!30, opacity=1, font=\footnotesize},
		envy/.style={-Latex, dashed, black},
		choice/.style={-Latex, thick, green!60!black},
		cross/.style={->, thick, red}
		]
		
		\begin{scope}[xshift=0cm]
			\node[agent, fill=agentj] (i1) at (-1, 0) {$i$};
			\node[poolimg] (P1) at (1.45,-0.9)
			{\includegraphics[width=3.5cm]{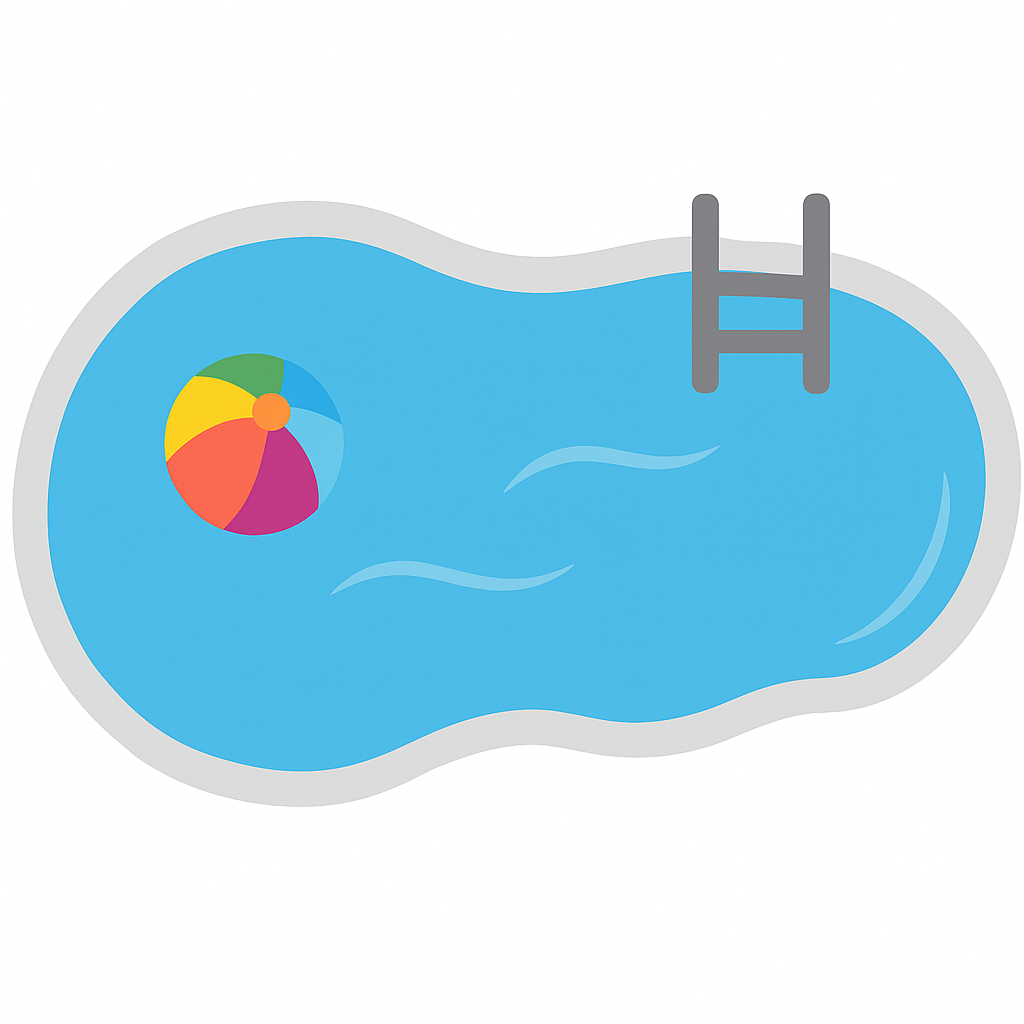}};
			\node[subpool] (SP1) at (1.5,-0.9) {{$\relevant{\pool}{i,j}$}};
			\node[bundle] (Bi1) at (-1,-1) {$\tiny\ralloc_i$};
			\draw[envy] (i1) to[bend right=30] (SP1);
			\draw (Bi1) to (i1);
			
		\end{scope}
		
		\begin{scope}[xshift=5.5cm]
			\node[agent, fill=agentj] (i2) at (-1, 0) {$i$};
			\node[agent] (j2) at (-1,-2) {$j$};
			\node[poolimg] (P2) at (1.45,-0.9)
			{\includegraphics[width=3.5cm]{figs/pool5}};
			\node[subpool] (SP2) at (1.5,-0.9) {};
			\node[bundle] (T2) at (SP2.center) {};
			\node[bundle] (Bi2) at (-1,-1) {$\tiny{\ralloc_i}$};
			\draw[envy] (i2) to[bend right=40] (T2);
						\draw (Bi2) to (i2);
		\end{scope}
		
		\begin{scope}[xshift=11.cm]
			\node[agent, fill=agentj] (i3) at (-1, 0) {$i$};
			\node[bundle] (T3) at (-1,-1) {};
			\node[poolimg] (P3) at (1.45,-0.9)
{\includegraphics[width=3.5cm]{figs/pool5}};
			\node[bundle] (Bi3) at (P3.center) {$\tiny{\ralloc_i}$};
			\draw[] (T3) -- (i3);
		\end{scope}
	\end{tikzpicture}}
\end{figure}
\vspace{-0.5cm}

\paragraph{Rule 2.} The second updating rule is designed to simplify the structure of $G_{0, \ralloc}$. This update rule applies if there exists an agent $i$ who has an out-degree of $0$ in $G_{0, \ralloc}$. By definition of $G_{0, \ralloc}$, this means that all other agents' bundles have value $0$ for agent $i$, and only bundle $\ralloc_i$ and $\pool$ might have nonzero value for agent $i$. We ask agent $i$ to pick the more valuable bundle from $\relevant{\pool}{i}$ and $\ralloc_i$. We allocate that bundle to agent $i$ and finalize her bundle. After this update, the utility of agent $i$ is more than the value of all remaining goods, and therefore $i$ does not envy the union of the remaining bundles (Property \textsf{(ii)}). Furthermore, since Rule 1 is not applicable, none of the remaining agents strongly envy agent $i$ (Property \textsf{(iv)}).

\begin{figure}[h]
	\centering
	\scalebox{0.7}{
	\begin{tikzpicture}[
		agent/.style={circle, draw, minimum size=0.9cm, font=\small, inner sep=1pt},
				subpool/.style={rectangle, draw=blue!70!black, thick,
			rounded corners, minimum width=0.9cm, minimum height=0.7cm,
			fill=white!30, opacity=1},
	bundle/.style={rectangle, draw=orange!70!black, rounded corners,
	minimum width=0.8cm, minimum height=0.6cm,
	fill=orange!30, opacity=1, font=\footnotesize},,
		final/.style={circle, draw, fill=green!20, minimum size=0.9cm, thick, inner sep=1pt},
		envy/.style={-Latex, thick, black},
		choice/.style={->, dashed},
		panelLabel/.style={font=\bfseries\footnotesize},
		checkmark/.style={green!60!black, font=\large, inner sep=0pt},
		cross/.style={red, font=\large, inner sep=0pt},
				poolimg/.style={inner sep=0pt, anchor=center, opacity=0.8},
		allocationbox/.style={  
			rounded rectangle,
			rounded corners=1pt,
			thick,
			dashed,
			inner sep=16mm,  
		}
		]
		\usetikzlibrary{positioning}
		
		\node[agent, fill=agentj] (i1) at (-2.4,0.9) {$i$};
		\node[agent] (j1) at (1.2, 0.9) {};
		\node[agent] (k1) at (1.2,-1.5) {};
		\node[agent] (l1) at (-0.6, 0.9) {};
		\node[agent] (m1) at (-0.6,-1.5) {};
		\node[agent] (n1) at (-2.4, -1.5) {};
		
		\draw[envy] (j1) to[bend left=10] (k1);
		\draw[envy] (l1) to[bend right=10] (m1);
		\draw[envy] (j1) to[bend right=10] (l1);
		\draw[envy] (m1) to[bend right=10] (j1);
		\draw[envy] (m1) to[bend left=10] (n1);
		\draw[envy] (n1) to[bend left=10] (i1);
		
		\node[poolimg] (P2) at (4.45,-0.9)
{\includegraphics[width=3.5cm]{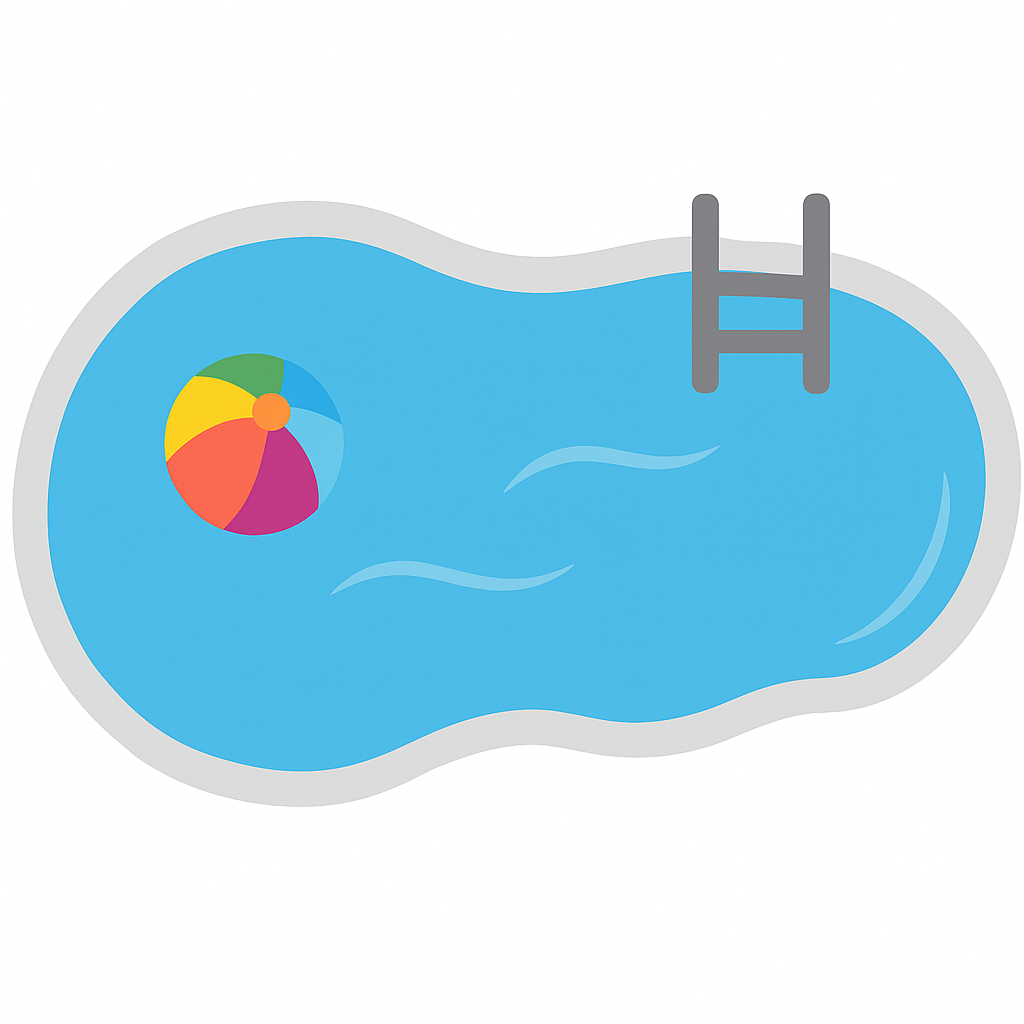}};

		\node[agent, fill=agentj] (i2) at (7,-0.3) {$i$};
		\node[bundle] (Mi2) at (4.5, 0.7) {$\ralloc_i$ \tiny};
		\node[subpool] (M0i2) at (4.5,-1.1) {$\relevant{\pool}{i}$ };
		
		
		\draw[choice] (Mi2) to[bend left=30] (i2);
		\draw[choice] (M0i2) to[bend right=30] (i2);
		
		\node[poolimg] (P2) at (10,-0.9)
{\includegraphics[width=3.5cm]{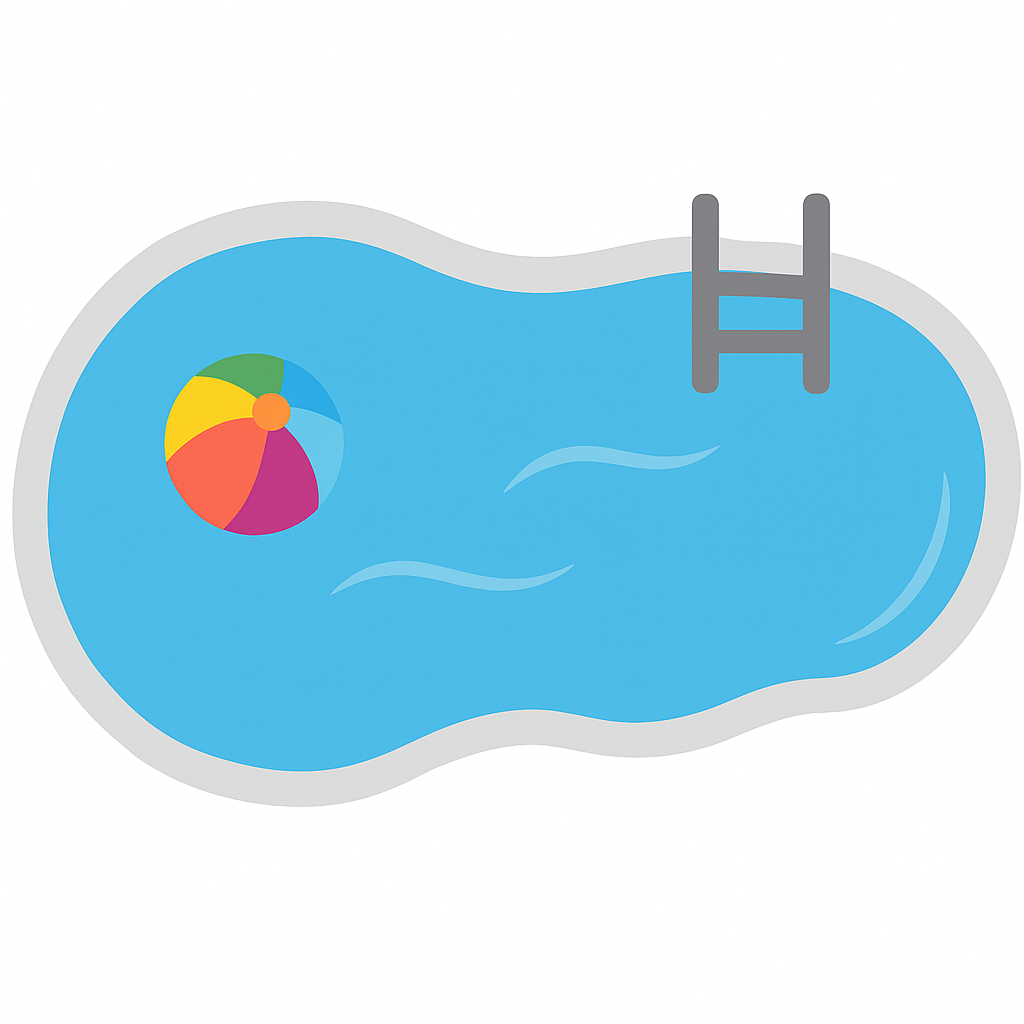}};
			\node[ darkgreen] at (10.5,-3) {Case 1: $i$ chooses $\relevant{\pool}{j}$};
		\node[final,fill=agenti] (i3) at (12.3,1) {$i$};
				\node[bundle] (Mi2) at (10, -1) {$\ralloc_i$ \tiny};
		\node[subpool, 
		below=0.4cm of i3, color=black, fill=green!20] (finalBundle) {$\relevant{\pool}{i}$};
		\draw (finalBundle) to[thick] (i3);

		
				\node[poolimg] (P2) at (15,-0.9)
		{\includegraphics[width=3.5cm]{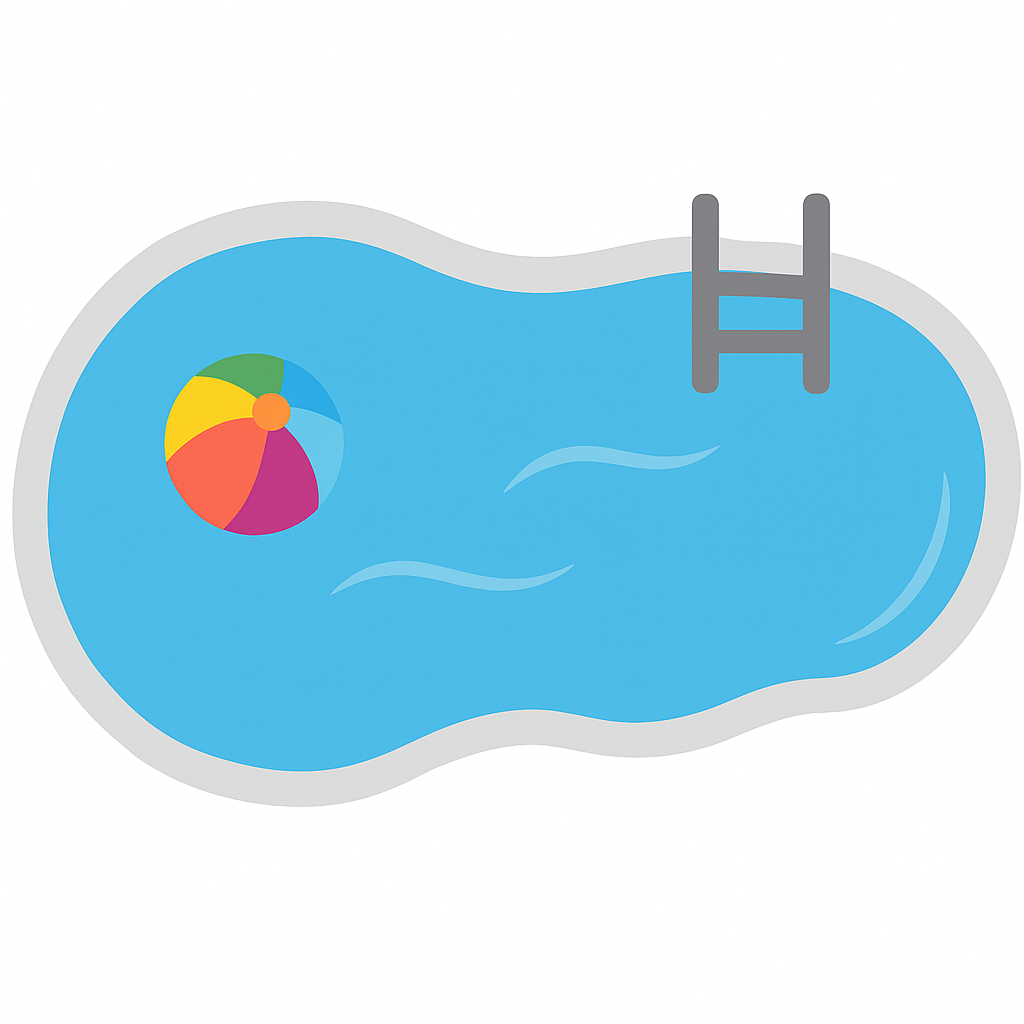}};
		\node[ darkgreen] at (15.5,-3) {Case 2: $i$ chooses $\ralloc_i$};
		\node[final,fill=agenti] (i3) at (17.3,1) {$i$};
		\node[subpool] (Mi2) at (15, -1) {$\relevant{\pool}{i}$ \tiny};
		\node[subpool, 
		below=0.4cm of i3, color=black, fill=green!20] (finalBundle) {$\ralloc_{i}$};
		\draw (finalBundle) to[thick] (i3);
		
	\end{tikzpicture}
}
\end{figure}

Recall that by Property $\textsf{(v)}$, every vertex in $G_{0,\ralloc}$ has in-degree at most $1$. When the first two rules are not applicable, no agent with out-degree zero exists in $G_{0, \ralloc}$, and hence, $G_{0, \ralloc}$ is consisted of disjoint cycles. The next three rules are designed to resolve cycles in the allocation graph.

\paragraph{Rule 3.} The most intricate case arises when dealing with cycles of length 2, which we resolve using the third updating rule. In this rule, we leverage \cref{thm:two} from \Citet{mahara2022twovaluations}, which ensures that any two-agent instance with a partial \(\efx\) allocation can be transformed into a complete \(\efx\) allocation without decreasing either agent's utility. Using this result, if the two agents involved are the only ones remaining, we apply \cref{thm:two} to allocate the remaining goods and finalize the allocation.

Otherwise, suppose these agents are $i$ and $j$ and one of the agents--say, agent $j$--does not envy the other. Our algorithm considers three bundles: (1) $\ralloc_i \cup \relevant{\pool}{i} \setminus \relevant{\pool}{i,j}$, (2) $\ralloc_j$, and (3) $\relevant{\pool}{j}$. Agent $j$ selects the more valuable bundle between (2) and (3). If $j$ chooses $\ralloc_j$, the algorithm allocate bundle (1) to $i$ and finalizes both agents. Otherwise, it gives (3) to $j$, and lets $i$ choose between (1) and (2). The unchosen bundle is returned to the pool, and both agents are marked as satisfied.

\begin{figure}[h!]
	\centering
	\scalebox{0.8}{
	\begin{tikzpicture}[
		agent/.style={circle, draw, minimum size=8mm, font=\small, inner sep=1pt},
		bundle/.style={rectangle, draw=orange!70!black, rounded corners,
	minimum width=0.65cm, minimum height=0.45cm,
	fill=orange!30, opacity=1, font=\footnotesize},						subpool/.style={rectangle, draw=blue!70!black, thick,
	rounded corners, minimum width=1.1cm, minimum height=0.7cm,
	fill=white!30, opacity=0.5},
			smallset/.style={rectangle, draw=green!70!black, thick,
			rounded corners, minimum width=1.2cm, minimum height=0.6cm,
			fill=green!20, opacity=1},
		envy/.style={-{Stealth[length=2mm]}, thick, black},
		return/.style={-{Stealth[length=2mm]}, dashed, red},
		swap/.style={<->, thick, magenta},
		poolimg/.style={inner sep=0pt, anchor=center, opacity=0.8, minimum width=3.5cm, minimum height=1.5cm}
		]
		
\begin{scope}[xshift=-5cm]
			\node[agent, fill=agentj] (i) at (0,0) {$i$};
	\node[agent, fill=agentj] (j) at (2,0) {$j$};
	
	\node[bundle] (A) at (0,-1) 
	{$\ralloc_i $};
	
	\node[bundle] (B) at (2,-1) 
	{$\ralloc_j$};

	\draw[envy,bend left=30](i) to (j);
	\draw[envy,bend left=30](j) to (i);
	\end{scope}
		\begin{scope}[xshift=0cm, yshift=0cm]
			\node[agent, fill=agentj] (i) at (0,0) {$i$};
			\node[agent, fill=agentj] (j) at (2,0) {$j$};
			
			\node[bundle] (A) at (0,-1) 
			{$\ralloc_i $};
			
			\node[bundle] (B) at (2,-1) 
			{$\ralloc_j$};


			\begin{scope}[on background layer]
				\node[poolimg] (P2) at (1,-3) {\includegraphics[width=3.5cm]{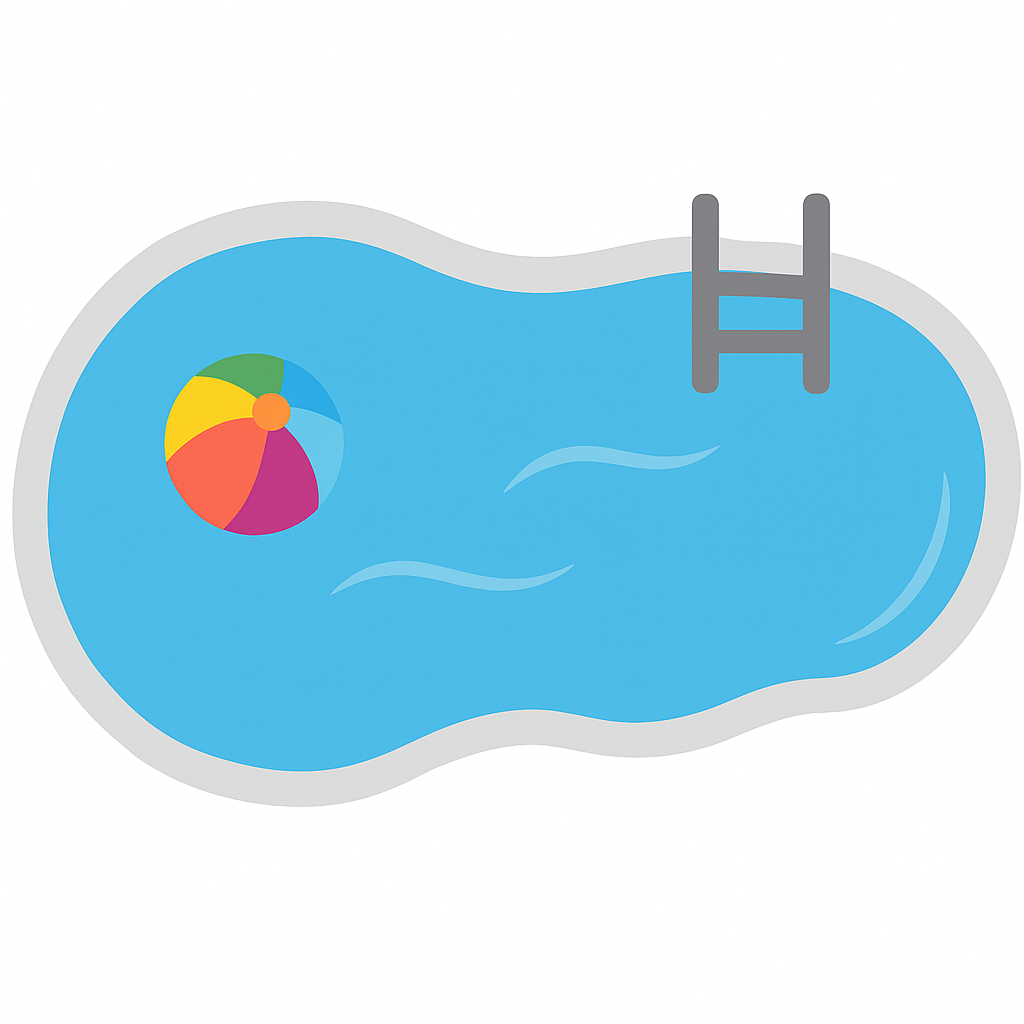}};
			\end{scope}
			
			\node[subpool] (Ri) at (1.3,-2.8) {\tiny$\relevant{\pool}{i}\setminus \relevant{\pool}{j}$};
			\node[subpool] (Rj) at (0.5,-3.2) {\tiny$\relevant{\pool}{j}$};
			\node[rectangle, draw=red, dotted, rounded corners,fill=crimsonglory,fill opacity=0.15,minimum height=0.85cm,minimum width=0.85cm]  (r1) at (B.center) {};
			\draw[dotted, rounded corners,fill=crimsonglory,fill opacity=0.15];
			
			\node[rectangle, draw=red, dotted, rounded corners,fill=crimsonglory,fill opacity=0.15,minimum height=0.83cm,minimum width=1.23cm]  (r2) at (Rj.center) {};
			\draw[dotted, red, rounded corners,fill=crimsonglory,fill opacity=0.15]
 %
 ($(A.north west)+(-0.2,0.2)$) --
 %
 ($(A.north east)+(0.2,0.2)$) --
 %
 ($(Ri.north east)+(0.1,0.1)$) --
 %
 ($(Ri.south east)+(0.1,-0.1)$) --
 %
 ($(Rj.east)+(0.1,-0.08)$) --
  ($(Rj.north east)+(0.1,0.09)$) --
  ($(Ri.north west)+(-0.2,-0.3)$) --
 %
 ($(A.south west)+(-0.2,-0.2)$) -- cycle;
 \draw[choice,bend left=40] (0.3,-2.8) to (j){};
 \draw[choice,bend right=40] (B.east) to (j.east){};
		\end{scope}
		
		\begin{scope}[xshift=5cm, yshift=0cm]
			\node[font=\footnotesize, darkgreen] at (1.5,-4.5) {Case 1: $j$ chooses $\ralloc_j$};
			
			\node[agent, fill=agenti] (i) at (0,0) {$i$};
			\node[agent, fill=agenti] (j) at (2,0) {$j$};
			
			\node[subpool, fill=white,draw=black,thin, fill opacity=0,text opacity=1,draw opacity=1,dotted] (Ri) at (1.3,-2.8) {\tiny$\relevant{\pool}{i}\setminus \relevant{\pool}{j}$};
			\node[subpool,opacity=1] (Rj) at (0.5,-3.2) {\tiny$\relevant{\pool}{j}$};
			
			\node[bundle, draw=black, dotted, fill=green!20] (A) at (0,-1) 
			{$\ralloc_i$};
			
			\node[bundle, draw=black, thick,fill=green!20] (B) at (2,-1.0) 
			{$\ralloc_j$};

			\begin{scope}[on background layer]
				\node[poolimg] (P3) at (1.1,-3) {\includegraphics[width=3.5cm]{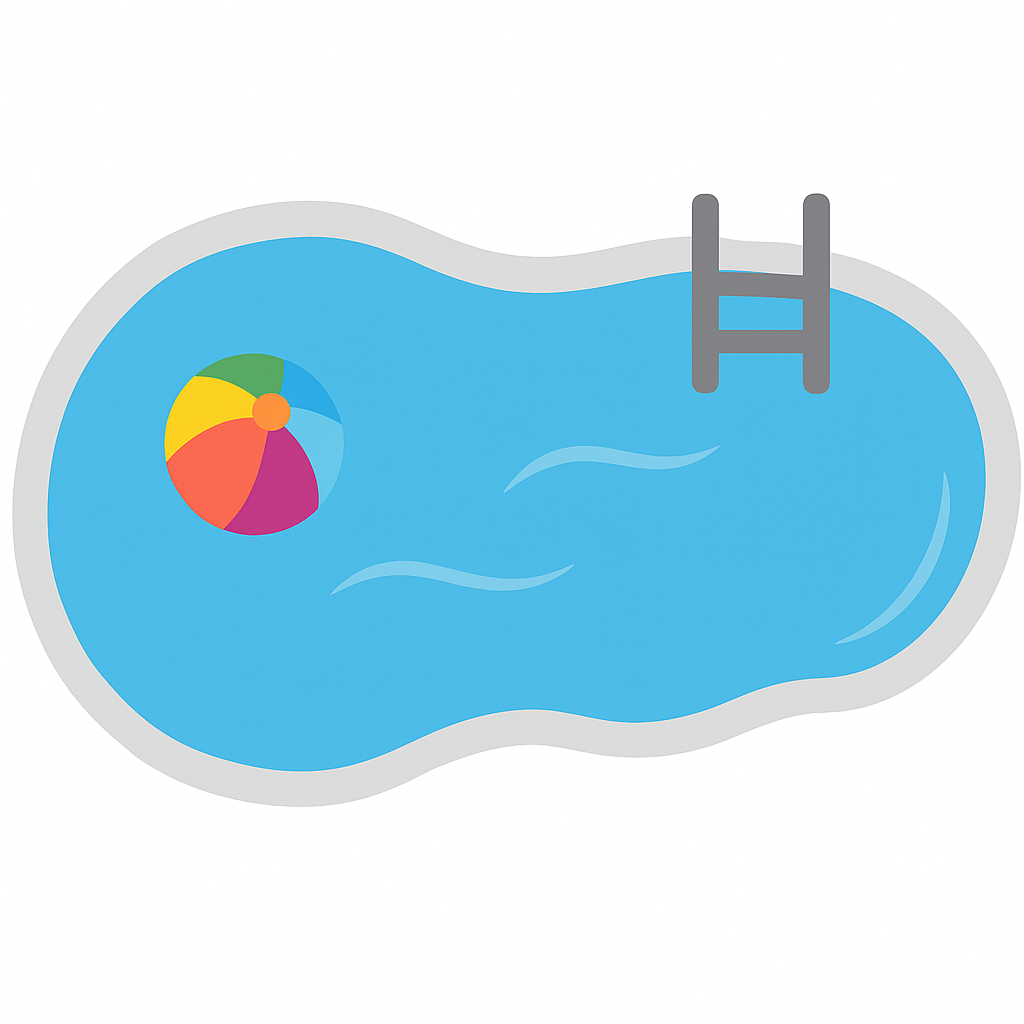}};
				\draw[dotted, dashed,black,thick, rounded corners,fill=green!20,fill opacity=1]
				%
				($(A.north west)+(-0.2,0.2)$) --
				%
				($(A.north east)+(0.2,0.2)$) --
				%
				($(Ri.north east)+(0.1,0.1)$) --
				%
				($(Ri.south east)+(0.1,-0.1)$) --
				%
				($(Rj.east)+(0.01,-0.01)$) --
				($(Rj.north east)+(0.01,0.01)$) --
				($(Ri.north west)+(-0.01,-0.4)$) --
				($(A.south west)+(-0.2,-0.2)$) -- cycle;
				\draw (0,-0.5) -- (i);
				\draw (B) -- (j);
				
			\end{scope}
		\end{scope}
		
		\begin{scope}[xshift=-5cm, yshift=-5.5cm]
			\node[font=\footnotesize, darkgreen] at (1.5,-4.5) {Case 2: $j$ chooses $\relevant{\pool}{j}$};
			
			\node[agent, fill=agenti] (i) at (0,0) {$i$};
			\node[agent, fill=agenti] (j) at (2,0) {$j$};

			
			\node[bundle, draw=black, thick,fill=green!20] (B) at (2,-1.0) 
{$\relevant{\pool}{j}$};
			\node[bundle] (A) at (0,-1) 
{$\ralloc_i $};
			\node[subpool,opacity=1] (Rj) at (0.5,-3.2) {\tiny$\ralloc_j$};			
						\node[subpool,minimum width=0.4] (Ri) at (1.7,-2.8) {\tiny$\relevant{\pool}{i}\setminus \relevant{\pool}{j}$};
						\node[rectangle, draw=red, dotted, rounded corners,fill=crimsonglory,fill opacity=0.15,minimum height=0.83cm,minimum width=1.23cm]  (r2) at (Rj.center) {};
						\draw[dotted, red, rounded corners,fill=crimsonglory,fill opacity=0.15]
			%
			($(A.north west)+(-0.2,0.2)$) --
			%
			($(A.north east)+(0.2,0.2)$) --
			%
			($(Ri.north east)+(0.1,0.1)$) --
			%
			($(Ri.south east)+(0.1,-0.1)$) --
			%
			($(Rj.east)+(0.1,-0.08)$) --
			($(Ri.north west)+(-0.2,-0.3)$) --
			%
			($(A.south west)+(-0.2,-0.2)$) -- cycle;
						\draw (B) -- (j);
\begin{scope}[on background layer]
	\node[poolimg] (P4) at (1,-3) {\includegraphics[width=3.5cm]{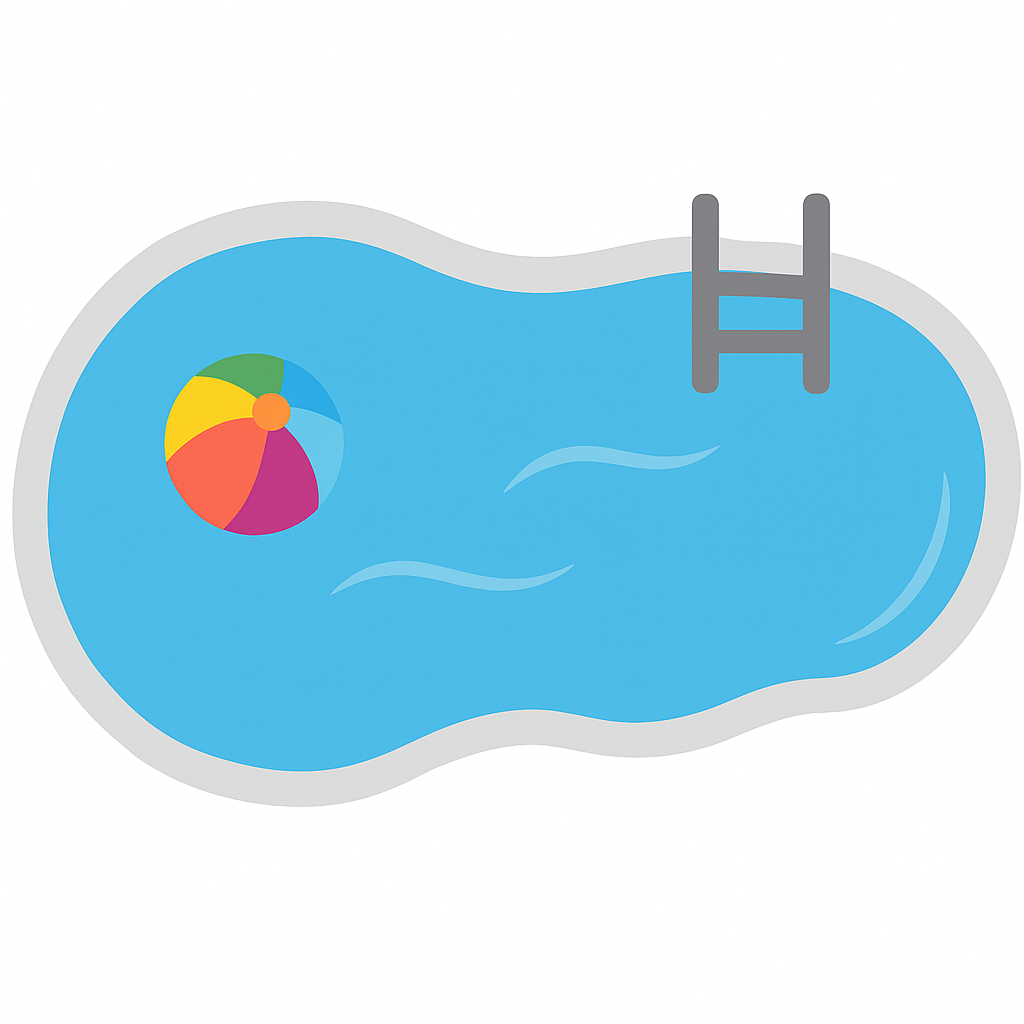}};
\end{scope}
			 \draw[choice,bend right=40] (1,-1) to (i){};
			\draw[choice,bend left=40] (Rj.west) to (i.west){};
		\end{scope}

		\begin{scope}[xshift=0cm, yshift=-5.5cm]
			\node[font=\footnotesize, darkgreen] at (1.5,-4.5) {Case 2.1: $i$ chooses $\ralloc_i \cup \relevant{\pool}{i} \setminus \relevant{\pool}{j}$};
			
			\node[agent, fill=agenti] (i) at (0,0) {$i$};
			\node[agent, fill=agenti] (j) at (2,0) {$j$};
			
			\node[subpool, fill=white,draw=black,thin, fill opacity=0,text opacity=1,draw opacity=1,dotted] (Ri) at (1.3,-2.8) {\tiny$\relevant{\pool}{i}\setminus \relevant{\pool}{j}$};
			\node[subpool,opacity=1] (Rj) at (0.5,-3.2) {\tiny$\ralloc_{j}$};
			
			\node[bundle, draw=black, dotted, fill=green!20] (A) at (0,-1) 
			{$\ralloc_i$};
			
			\node[bundle, draw=black, thick,fill=green!20] (B) at (2,-1.0) 
			{$\relevant{\pool}{j}$};

			\begin{scope}[on background layer]
				\node[poolimg] (P3) at (1.1,-3) {\includegraphics[width=3.5cm]{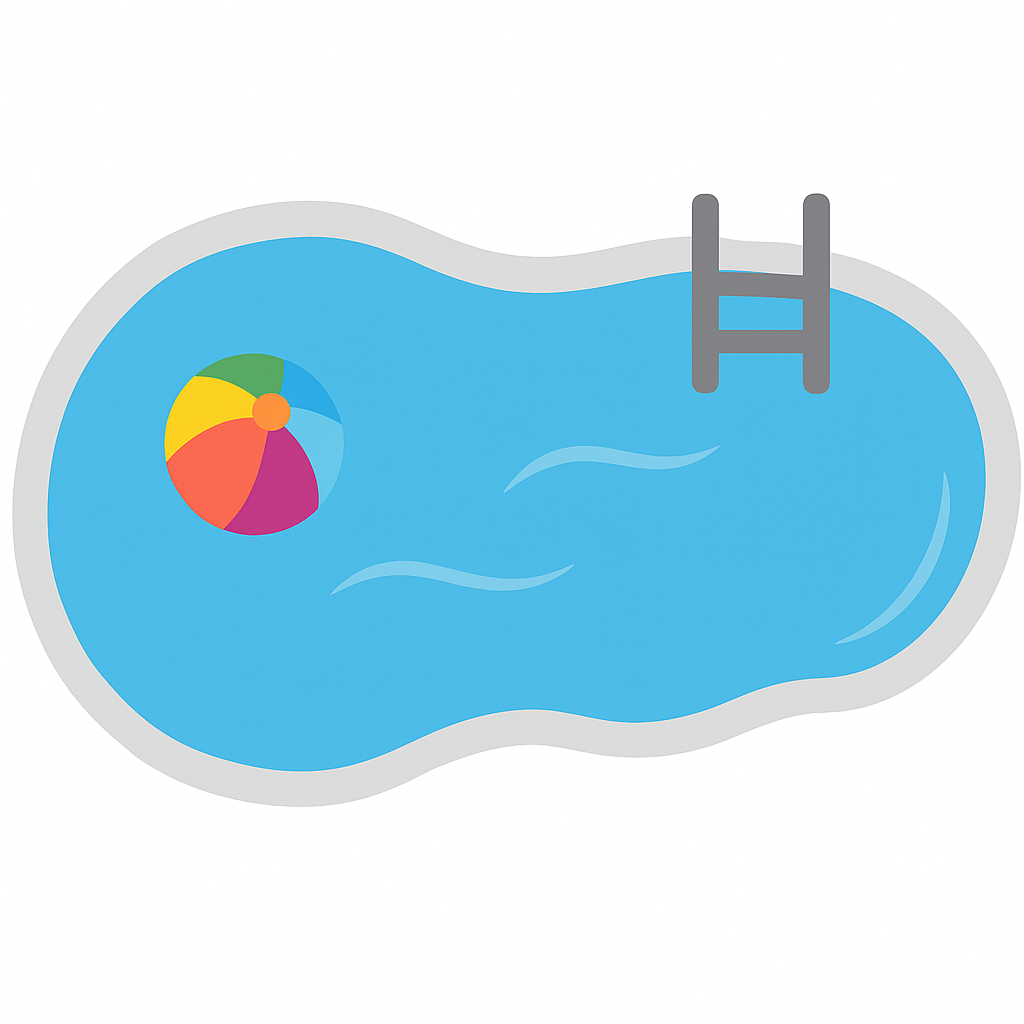}};
				\draw[dotted, dashed,black,thick, rounded corners,fill=green!20,fill opacity=1]
				%
				($(A.north west)+(-0.2,0.2)$) --
				%
				($(A.north east)+(0.2,0.2)$) --
				%
				($(Ri.north east)+(0.1,0.1)$) --
				%
				($(Ri.south east)+(0.1,-0.1)$) --
				%
				($(Rj.east)+(0.01,-0.01)$) --
				($(Rj.north east)+(0.01,0.01)$) --
				($(Ri.north west)+(-0.01,-0.4)$) --
				($(A.south west)+(-0.2,-0.2)$) -- cycle;
				\draw (0,-0.5) -- (i);
				\draw (B) -- (j);
				
			\end{scope}
		\end{scope}
		
		\begin{scope}[xshift=5cm, yshift=-5.5cm]
			\node[font=\footnotesize, darkgreen] at (1.5,-4.5) {Case 2.2: $i$ chooses $\ralloc_{j}$};
			
			\node[agent, fill=agenti] (i) at (0,0) {$i$};
			\node[agent, fill=agenti] (j) at (2,0) {$j$};
			
			
			\node[bundle, draw=black, thick,fill=green!20] (B) at (2,-1.0) 
			{$\relevant{\pool}{j}$};
			\node[bundle,fill=green!20,draw=black,thick] (A) at (0,-1) 
			{$\ralloc_j $};
			\node[subpool,opacity=1] (Rj) at (0.5,-3.4) {\tiny$\ralloc_i$};			
			\node[subpool,minimum width=0.4,opacity=1] (Ri) at (1.7,-2.8) {\tiny$\relevant{\pool}{i}\setminus \relevant{\pool}{j}$};
			\draw (B) -- (j);
			\begin{scope}[on background layer]
				\node[poolimg] (P4) at (1,-3) {\includegraphics[width=3.5cm]{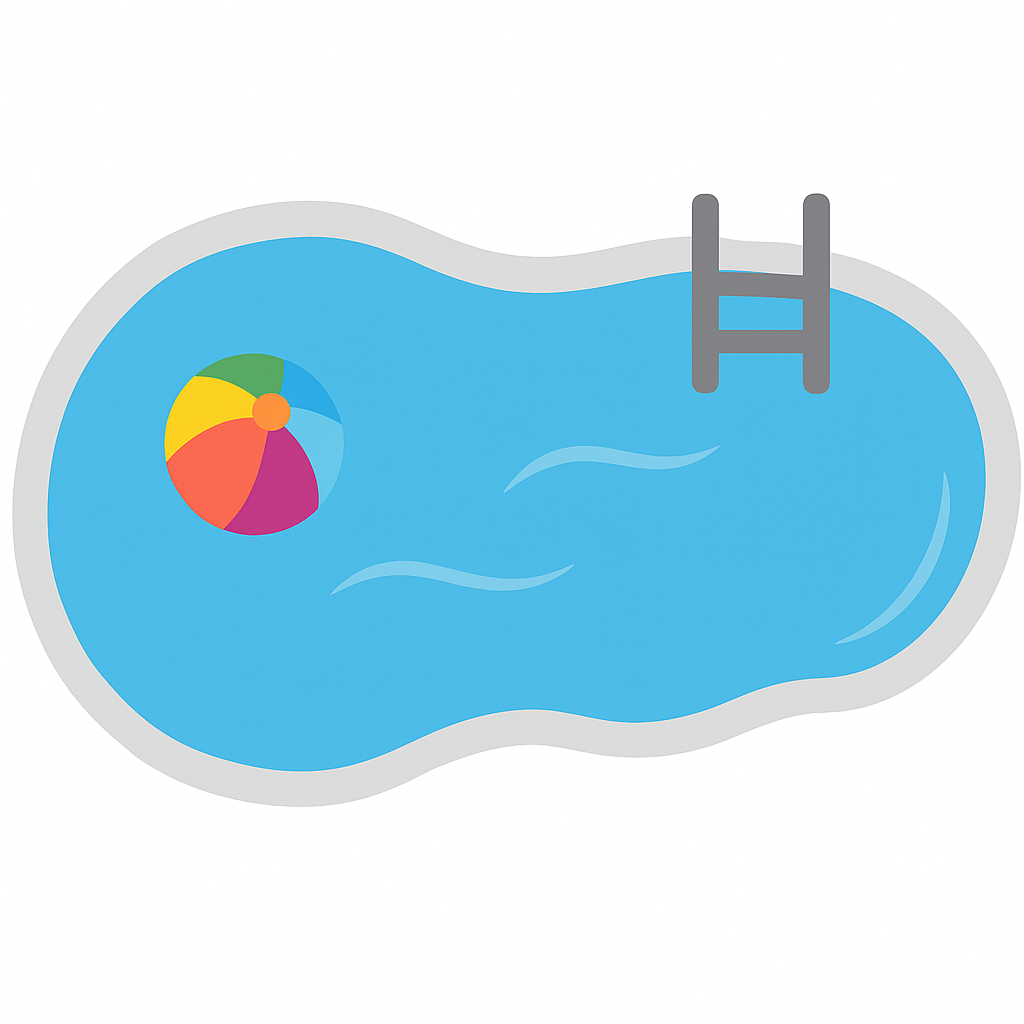}};
			\end{scope}
\draw (i) -- (A);
		\end{scope}

	\end{tikzpicture}}
	\label{fig:rule3}
\end{figure}
\vspace{-0.5cm}
Once all 2-cycles have been resolved, the algorithm proceeds to handle the remaining cycles using the final two update rules. These cycles are classified as either \emph{homogeneous} or \emph{heterogeneous}. We call an edge \emph{heavy} if its weight is greater than \(\nicefrac{1}{\sqrt{2}}\), and \emph{light} otherwise. A cycle is \emph{homogeneous} if all of its edges are either heavy or all light; otherwise, it is heterogeneous. Rule~4 is responsible for resolving homogeneous cycles, while Rule~5 is used for heterogeneous ones.

\paragraph{Rule 4.} Rule~4 applies to a homogeneous cycle \(C\). If the cycle consists of heavy edges, we perform a \emph{rotation}: each agent in the cycle receives the bundle of their successor. This update may reduce each agent's value, but only by a factor of at most \(\sqrt{2}\), since the edge weights in the original cycle are all above \(\nicefrac{1}{\sqrt{2}}\). Moreover, by construction of  \(G_{0,\ralloc}\), this rotation reverses the direction of the edges in \(C\). If the cycle is light, we skip the rotation and retain the current allocation.

Next, we allocate additional goods from the pool that are relevant to the agents in the cycle. For each such good \(g\), we allocate it to an agent \(i \in C\) who finds \(g\) relevant, but whose predecessor in the cycle does not. Since the cycle has length greater than two, such an agent always exists. This step ensures that after the update, no agent in the cycle envies their successor by more than a factor of \(\sqrt{2}\). Finally, all agents in the cycle are moved to set of satisfied agents.

\begin{figure}[h!]
	\centering
	\scalebox{0.8}{
	\begin{tikzpicture}[
		agentj/.style={circle, draw, minimum size=7mm, font=\small, inner sep=1pt, fill=agentj}, 
		agenti/.style={circle, draw, minimum size=7mm, font=\small, inner sep=1pt, fill=agenti}, 
		bundle/.style={rectangle, draw=orange!70!black, rounded corners,
			minimum width=0.7cm, minimum height=0.5cm,
			fill=orange!30, opacity=1, font=\scriptsize, align=center},
		heavy/.style={-{Stealth[length=2mm]}, thick, red, bend left=15},
		heavyrev/.style={-{Stealth[length=2mm]}, thick, red, bend right=15}, 
		light/.style={-{Stealth[length=2mm]}, thick, green!70!black, bend left=15},
		rotation/.style={-{Stealth[length=2mm]}, thick, magenta},
		choice/.style={-{Stealth[length=2mm]}, cyan},
		subpool/.style={rectangle, draw=blue!70!black, thick,
			rounded corners, minimum width=1cm, minimum height=0.6cm,
			fill=white, opacity=1, font=\scriptsize, align=center},
		smallset/.style={rectangle, draw=green!70!black, thick,
			rounded corners, minimum width=0.8cm, minimum height=0.4cm,
			fill=green!20, opacity=1},
		poolimg/.style={inner sep=0pt, anchor=center, opacity=0.8, minimum width=2.5cm, minimum height=1.0cm},
		rotind/.style={circle, draw=white, thick, minimum size=8mm}
		]
		
		\begin{scope}[xshift=-3cm]
			
			\node[agentj] (i1) at (0.5,2.5) {$i_1$};
			\node[agentj] (i2) at (1.5,3.5) {$i_2$};
			\node[agentj] (i3) at (2.5,2.5) {$i_3$};
			\node[agentj] (i4) at (1.5,1.5) {$i_4$}; 
			
			\draw[heavy] (i1) to (i2);
			\draw[heavy] (i2) to (i3);
			\draw[heavy] (i3) to (i4);
			\draw[heavy] (i4) to (i1);
			
			\node[rotind] at (1.5,2.5) {\Large $\circlearrowleft$};
			
			\node[bundle] (b1) at (-0.3,2.5) {$\ralloc_{i_1}$};
			\node[bundle] (b2) at (1.5,4.3) {$\ralloc_{i_2}$};
			\node[bundle] (b3) at (3.3,2.5) {$\ralloc_{i_3}$};
			\node[bundle] (b4) at (1.5,0.7) {$\ralloc_{i_4}$};
			\draw (b1) -- (i1);
			\draw (b2) -- (i2);
			\draw (b3) -- (i3);
			\draw (b4) -- (i4);
			
			\node[agentj] (j1) at (0.5,-2.0) {$j_1$};
			\node[agentj] (j2) at (1,-3.0) {$j_2$}; 
			\node[agentj] (j3) at (2,-3.0) {$j_3$};
			\node[agentj] (j4) at (2.5,-2.0) {$j_4$};
			\node[agentj] (j5) at (1.5,-1.0) {$j_5$}; 
			
			\draw[light] (j1) to (j2);
			\draw[light] (j2) to (j3);
			\draw[light] (j3) to (j4);
			\draw[light] (j4) to (j5);
			\draw[light] (j5) to (j1);
			
			\node[bundle] (b5) at (-0.3,-1.7) {$\ralloc_{j_1}$};
			\node[bundle] (b6) at (0.5,-3.7) {$\ralloc_{j_2}$};
			\node[bundle] (b7) at (2.5,-3.7) {$\ralloc_{j_3}$};
			\node[bundle] (b8) at (3.3,-1.7) {$\ralloc_{j_4}$};
			\node[bundle] (b9) at (1.5,-0.25) {$\ralloc_{j_5}$}; 
			\draw (b5) -- (j1);
			\draw (b6) -- (j2);
			\draw (b7) -- (j3);
			\draw (b8) -- (j4);
			\draw (b9) -- (j5);
			
		\end{scope}
		
\begin{scope}[xshift=2.5cm]
	
	\node[agentj] (i1) at (0.5,2.5) {$i_1$};
	\node[agentj] (i2) at (1.5,3.5) {$i_2$};
	\node[agentj] (i3) at (2.5,2.5) {$i_3$};
	\node[agentj] (i4) at (1.5,1.5) {$i_4$}; 
	
	\draw[light] (i2) to (i1);
	\draw[light] (i3) to (i2);
	\draw[light] (i4) to (i3);
	\draw[light] (i1) to (i4);
	
	\node[bundle] (b1) at (-0.3,2.5) {$\ralloc_{i_2}$};
	\node[bundle] (b2) at (1.5,4.3) {$\ralloc_{i_3}$};
	\node[bundle] (b3) at (3.3,2.5) {$\ralloc_{i_4}$};
	\node[bundle] (b4) at (1.5,0.7) {$\ralloc_{i_5}$};
	\draw (b1) -- (i1);
	\draw (b2) -- (i2);
	\draw (b3) -- (i3);
	\draw (b4) -- (i4);
	
	\node[agentj] (j1) at (0.5,-2.0) {$j_1$};
	\node[agentj] (j2) at (1,-3.0) {$j_2$}; 
	\node[agentj] (j3) at (2,-3.0) {$j_3$};
	\node[agentj] (j4) at (2.5,-2.0) {$j_4$};
	\node[agentj] (j5) at (1.5,-1.0) {$j_5$}; 
	
	\draw[light] (j1) to (j2);
	\draw[light] (j2) to (j3);
	\draw[light] (j3) to (j4);
	\draw[light] (j4) to (j5);
	\draw[light] (j5) to (j1);
	
	\node[bundle] (b5) at (-0.3,-1.7) {$\ralloc_{j_1}$};
	\node[bundle] (b6) at (0.5,-3.7) {$\ralloc_{j_2}$};
	\node[bundle] (b7) at (2.5,-3.7) {$\ralloc_{j_3}$};
	\node[bundle] (b8) at (3.3,-1.7) {$\ralloc_{j_4}$};
	\node[bundle] (b9) at (1.5,-0.25) {$\ralloc_{j_5}$}; 
	\draw (b5) -- (j1);
	\draw (b6) -- (j2);
	\draw (b7) -- (j3);
	\draw (b8) -- (j4);
	\draw (b9) -- (j5);
	
\end{scope}
		\begin{scope}[xshift=8cm,yshift=-1cm]
			
			\node[agenti] (i1) at (0.5,2.5) {$i_1$};
			\node[agenti] (i2) at (1.5,3.5) {$i_2$};
			\node[agenti] (i3) at (2.5,2.5) {$i_3$};
			\node[agenti] (i4) at (1.5,1.5) {$i_4$};
			
			\node[bundle,fill=green!20,draw=black,thick] (b1) at (-0.3,2.5) {$\ralloc_{i_2}$};
			\node[bundle,fill=green!20,draw=black,thick] (b2) at (1.5,4.3) {$\ralloc_{i_3}$};
			\node[bundle,fill=green!20,draw=black,thick] (b3) at (3.3,2.5) {$\ralloc_{i_4}$};
			\node[bundle,fill=green!20,draw=black,thick] (b4) at (1.5,0.7) {$\ralloc_{i_1}$};
			\draw (b1) -- (i1);
			\draw (b2) -- (i2);
			\draw (b3) -- (i3);
			\draw (b4) -- (i4);

			\begin{scope}[on background layer]
				\node[poolimg] (P3) at (1.5,-1.5) {\includegraphics[width=4.5cm]{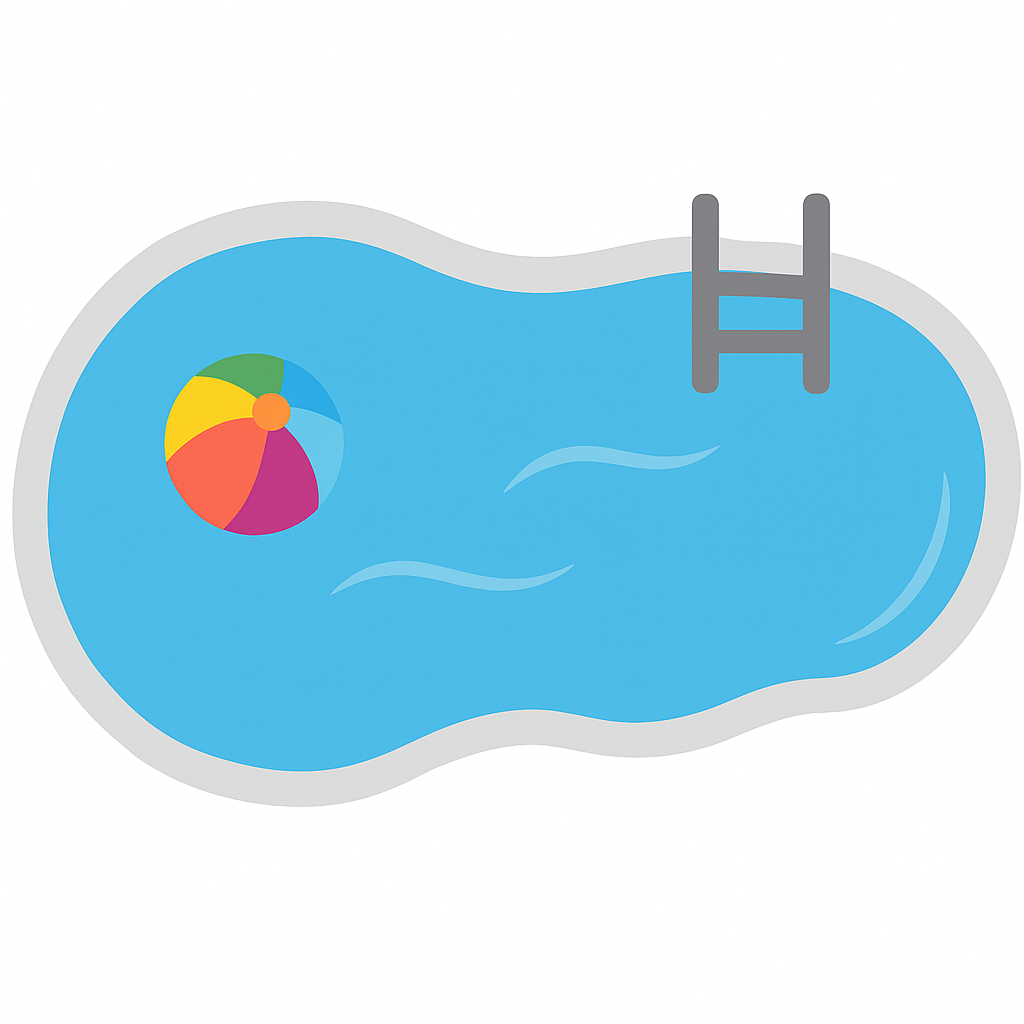}};
				\node[good,draw=black,fill=yellow!20] (Ri) at (2.6,-1.5) {\tiny$g_1 \in \relevant{\pool}{i_2,i_1}$}; 
				\node[good,draw=black,fill=yellow!20] (Rj) at (1.4,-1) {\tiny$g_2 \in \relevant{\pool}{i_1,i_3}$};
				\node[good,draw=black,fill=yellow!20] (Rk) at (0.3,-1.8) {\tiny$g_3 \in \relevant{\pool}{i_4}$};
			\end{scope}

			\draw[choice] (Ri) to[bend right=90] (b2.east);
			\draw[choice] (Rj) to[bend left=20] (b1.south);
			\draw[choice] (Rj) to[bend right=20] (b3.south);
			\draw[choice] (Rk) to (b4.south);
			\node[font=\scriptsize, right] at (1.5,-0.8) {};
			\node[font=\scriptsize, left] at (1.5,-2.2) {};
		\end{scope}
		
	\end{tikzpicture}}
	\label{fig:homogeneous_rule}
\end{figure}
\vspace{-0.5cm}

\paragraph{Rule 5.} The only remaining case is when the envy graph \( G_{0, \ralloc} \) consists entirely of heterogeneous cycles. Let \( C \) be such a cycle. By definition, there exist three consecutive agents \( i \rightarrow j \rightarrow k \) in \( G_{0, \ralloc} \) such that the edge \( i \rightarrow j \) is heavy, while \( j \rightarrow k \) is light. We perform the update based on these three agents as follows:
If agent \( j \)'s value for the pool satisfies \( \valu_j(\pool) \leq \valu_j(\ralloc_j)/\sqrt{2} \), we finalize \( j \)'s bundle, without any further changes. Otherwise, we allocate all goods in \( \relevant{\pool}{j} \) to agent $j$, and agent \( i \) receives the bundle \( \ralloc_j \cup \relevant{\pool}{i} \setminus \relevant{\pool}{i, j} \), and then we finalize both $i$ and $j$.

\begin{figure}[h!]
	\centering
	\scalebox{0.8}{
	\begin{tikzpicture}[
		agentj/.style={circle, draw, minimum size=7mm, font=\small, inner sep=1pt, fill=agentj}, 
		agenti/.style={circle, draw, minimum size=7mm, font=\small, inner sep=1pt, fill=agenti}, 
		bundle/.style={rectangle, draw=orange!70!black, rounded corners,
			minimum width=0.7cm, minimum height=0.5cm,
			fill=orange!30, opacity=1, font=\scriptsize, align=center},
		heavy/.style={-{Stealth[length=2mm]}, thick, red, bend left=15},
		heavyrev/.style={-{Stealth[length=2mm]}, thick, red, bend right=15}, 
		light/.style={-{Stealth[length=2mm]}, thick, green!70!black, bend left=15},
		rotation/.style={-{Stealth[length=2mm]}, thick, magenta},
		choice/.style={-{Stealth[length=2mm]}, cyan},
		subpool/.style={rectangle, draw=blue!70!black, thick,
			rounded corners, minimum width=1cm, minimum height=0.6cm,
			fill=white, opacity=1, font=\scriptsize, align=center},
		smallset/.style={rectangle, draw=green!70!black, thick,
			rounded corners, minimum width=0.8cm, minimum height=0.4cm,
			fill=green!20, opacity=1},
		poolimg/.style={inner sep=0pt, anchor=center, opacity=0.8, minimum width=2.5cm, minimum height=1.0cm},
		rotind/.style={circle, draw=white, thick, minimum size=8mm}
		]
		
		\begin{scope}[xshift=-3cm]
			\node[rectangle,rounded corners, dashed, minimum width=5cm,minimum height=2.75cm,draw=black,fill opacity=0.6] at (1.5,0.85){};
			\node[agentj] (j1) at (0.5,0) {$j_1$};
			\node[agentj] (j2) at (1,-1.0) {$j_2$};
			\node[agentj] (j3) at (2,-1.0) {$j_3$};
			\node[agentj] (j4) at (2.5,0) {$j_4$};
			\node[agentj] (j5) at (1.5,1.0) {$j_5$}; 
			
			\draw[light] (j1) to (j2);
			\draw[light] (j2) to (j3);
			\draw[light] (j3) to (j4);
			\draw[heavy] (j4) to (j5);
			\draw[light] (j5) to (j1);

			\node[bundle] (b5) at (-0.3,0) {$\ralloc_{j_1}$};
			\node[bundle] (b6) at (0.5,-1.7) {$\ralloc_{j_2}$};
			\node[bundle] (b7) at (2.5,-1.7) {$\ralloc_{j_3}$};
			\node[bundle] (b8) at (3.3,0) {$\ralloc_{j_4}$};
			\node[bundle] (b9) at (1.5,1.75) {$\ralloc_{j_5}$}; 
			\draw (b5) -- (j1);
			\draw (b6) -- (j2);
			\draw (b7) -- (j3);
			\draw (b8) -- (j4);
			\draw (b9) -- (j5);
									\node[ darkgreen] at (7,-5.5) {Case 1: \( \valu_j(\pool) \leq \valu_j(\ralloc_j)/\sqrt{2} \)};
			
		\end{scope}
		
		\begin{scope}[xshift=2.5cm]
			\node[agentj] (j1) at (0.5,0) {$j_1$};
			\node[agentj] (j2) at (1,-1.0) {$j_2$};
			\node[agentj] (j3) at (2,-1.0) {$j_3$};
			\node[agentj] (j4) at (2.5,0) {$j_4$};
			\node[agenti] (j5) at (1.5,1.0) {$j_5$}; 
			
			\draw[light] (j1) to (j2);
			\draw[light] (j2) to (j3);
			\draw[light] (j3) to (j4);

			\node[bundle] (b5) at (-0.3,0) {$\ralloc_{j_1}$};
			\node[bundle] (b6) at (0.5,-1.7) {$\ralloc_{j_2}$};
			\node[bundle] (b7) at (2.5,-1.7) {$\ralloc_{j_3}$};
			\node[bundle] (b8) at (3.3,0) {$\ralloc_{j_4}$};
			\node[bundle,fill=green!20,draw=black,thick] (b9) at (1.5,1.75) {$\ralloc_{j5}$}; 
			\draw (b5) -- (j1);
			\draw (b6) -- (j2);
			\draw (b7) -- (j3);
			\draw (b8) -- (j4);
			\draw (b9) -- (j5);
			
						\node[ darkgreen] at (7,-5.5) {Case 2: \( \valu_j(\pool) > \valu_j(\ralloc_j)/\sqrt{2} \)};
			
		\end{scope}
		\begin{scope}[xshift=8cm]
\node[agentj] (j1) at (0.5,2) {$j_1$};
\node[agentj] (j2) at (1,1.0) {$j_2$};
\node[agentj] (j3) at (2,1.0) {$j_3$};
\node[agenti] (j4) at (3,-1) {$j_4$};
\node[agenti] (j5) at (0,-1.0) {$j_5$}; 

\draw[light] (j1) to (j2);
\draw[light] (j2) to (j3);

\node[bundle] (b5) at (-0.3,2) {$\ralloc_{j1}$};
\node[bundle] (b6) at (0.5,0.3) {$\ralloc_{j2}$};
\node[bundle] (b7) at (2.5,0.3) {$\ralloc_{j3}$};
\node[bundle] (b8) at (2.8,-3.5) {\tiny$\ralloc_{j4}$};
\node[bundle,fill=green!20] (b9) at (1.5,-1) {$\ralloc_{j5}$}; 
\draw (b5) -- (j1);
\draw (b6) -- (j2);
\draw (b7) -- (j3);

\begin{scope}[on background layer]
	\node[poolimg] (P2) at (1.3,-3.5) {\includegraphics[width=4.5cm]{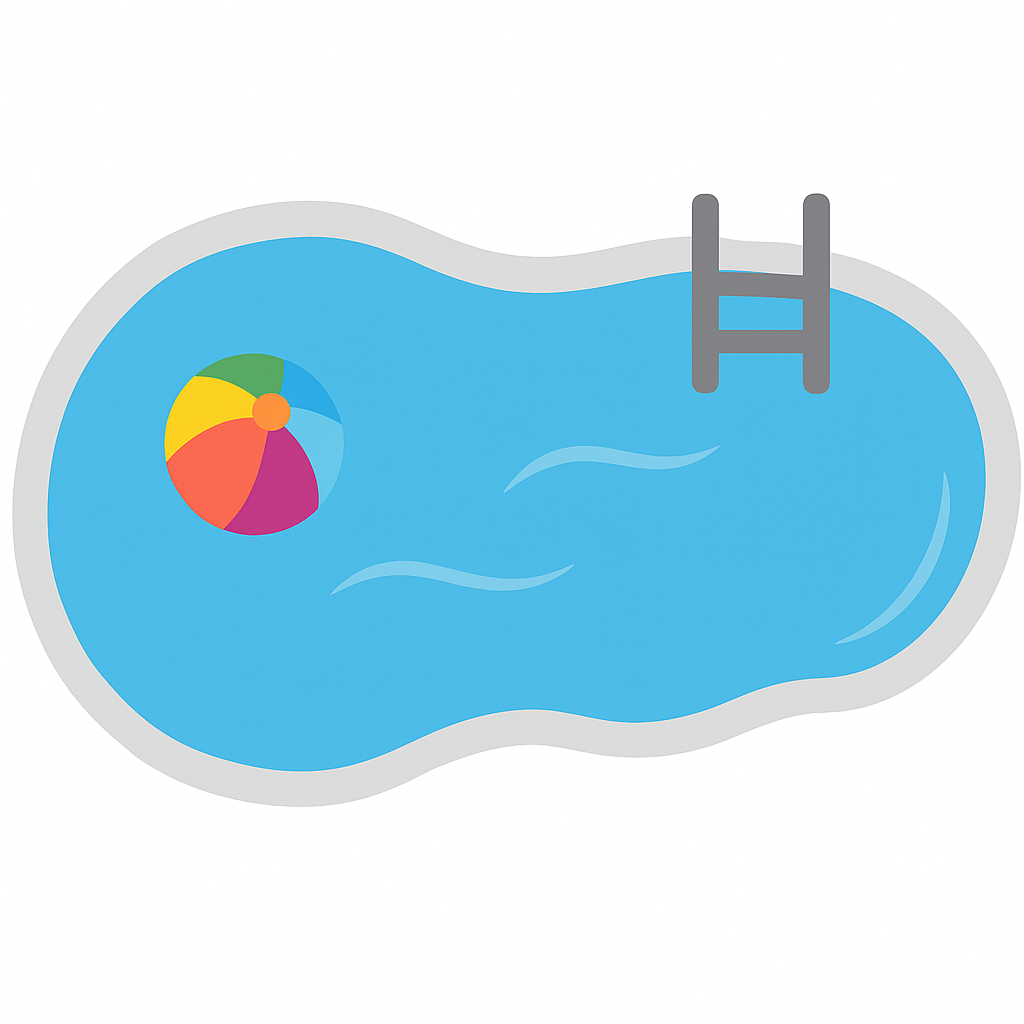}};
\end{scope}

\draw[choice,bend left=30] (b8) to (3,-3);

\node[subpool] (Ri) at (1.6,-3.4) {\tiny$\relevant{\pool}{i}$};
\node[subpool,fill=green!20,draw=black] (Rj) at (0.8,-3.8) {\tiny$\relevant{\pool}{j}$};

			\draw [dashed, thick, black, rounded corners,fill=green!20,fill opacity=0.5]
%
($(b9.north west)+(-0.2,0.2)$) --
%
($(b9.north east)+(0.2,0.2)$) --
%
($(Ri.north east)+(0.1,0.1)$) --
%
($(Ri.south east)+(0.1,-0.1)$) --
%
($(Rj.east)+(0.1,-0.08)$) --
($(Rj.north east)+(0.1,0.09)$) --
($(Ri.north west)+(-0.2,-0.3)$) --
%
($(b9.south west)+(-0.2,-0.2)$) -- cycle;

\draw[bend left=30] (Rj) to (j5);
\draw[bend right=30] (2.2,-2.5) to (j4);
		\end{scope}
		
	\end{tikzpicture}}

	\label{fig:heterogenous_rule}
\end{figure}

\paragraph{Final step.} When none of the rules are applicable, the bundles of all agents are finalized, but some goods may still remain in the pool. In the final step, we allocate these goods to the last agent whose bundle was finalized and return the allocation. 
The final allocation is ---by properties $(\textsf{i})$ and $(\textsf{ii})$--- $(\nicefrac{1}{\sqrt{2}})$-$\efx$.

	\section{$(\nicefrac{1}{\sqrt{2}})$-$\efx$ Allocation Algorithm}\label{sec:p2}

In this section, we outline our algorithm in more detail. Algorithm~\ref{alg1} shows a pseudocode of our method. We describe the updating rules one by one, and for each rule, we prove that it preserves properties~$\textsf{(i)}$ to~$\textsf{(v)}$.
These properties are trivially satisfied at the beginning of the algorithm.

\begin{algorithm}[h]
	\caption{$(\nicefrac{1}{\sqrt 2})$-$\efx$ Allocation Algorithm}
	\label{alg1}
	\textbf{Initialize: }$\salloc =(), \sagents = \emptyset, \ragents = \agents, \ralloc = $ basic feasible allocation.\;
	\While{there exists an applicable update rule}{
		\For{$k \gets 1$ \KwTo $5$}{
			\If{Rule $k$ is applicable}{
				Apply Rule $k$\;
				\textbf{break}\;
			}
		}
	}
	Apply the final step\;
	\Return{$\salloc$}\;
\end{algorithm}
\vspace{-0.5cm}

\subsection*{Update Rule 1}\label{Rule:1}
The structure of this rule was previously outlined in Section~\ref{sec:resultstech}. This rule is applicable when there are two agents, $i$ and $j$, where at least one of them envies $\relevant{\pool}{i, j}$. If so, we identify the minimal subset, $S \subseteq \relevant{\pool}{i, j}$ that is envied by either $i$ or $j$. Assume, without loss of generality, that agent $i$ envies $S$ (can be $j$ too). We allocate $S$ to agent $i$ and return her previous bundle, $\ralloc_i$, to the pool. Algorithm~\ref{alg:rule1} provides the pseudocode for this updating rule.

\vspace{0.5cm}
\begin{algorithm}[H]
	\caption{Update Rule 1}
	\label{alg:rule1}
	\textbf{Input:} \(\ralloc,\ragents,\salloc,\sagents,\pool\)\;
	\If{\(\nexists\) pair \((i,j)\in \ragents\) with \(\valu_i(\ralloc_i) < \valu_i(\relevant{\pool}{i,j})\)}{
		\Return (\textit{Not applicable})\;
	}
	Let \((i,j)\) be a pair with \(\valu_i(\ralloc_i) < \valu_i(\relevant{\pool}{i,j})\)\;
	\(\mathcal{T} \gets \{\,T\subseteq \relevant{\pool}{i,j}\mid i\text{ or }j\text{ envy }T\}\)\;
	\(S \gets \arg\min_{T\in\mathcal{T}}|T|\)\;
	\If{\(i\) envies \(S\)}{
		return \(\ralloc_i\) to \(\pool\)\;
		\(\ralloc_i \gets S\)\;
	}\Else{
		return \(\ralloc_j\) to \(\pool\)\;
		\(\ralloc_j \gets S\)\;
	}
\end{algorithm}

\begin{lemma}\label{lem:rule1}
	Rule 1 preserves properties $\textsf{(i)}$ to $\textsf{(v)}$.
\end{lemma}

\begin{proof}
	Let \( i \) be the agent who receives the new bundle \( S \subseteq \relevant{\pool}{i,j} \). We verify that each property remains valid after the update:
	
	\begin{itemize}
		\item[\textsf{(i)}] This property pertains to the final allocation. Since Rule 1 only modifies the allocation of a remaining agent and does not finalize any agent, it has no effect on the final allocation.
		
		\item[\textsf{(ii)}] This property continues to hold trivially after the update.
		
		\item[\textsf{(iii)}] By construction, agent \( j \) does not strongly envy \( i \)'s new bundle. Moreover, all other remaining agents assign zero value to the new bundle of \( i \), so this property is preserved.
		
		\item[\textsf{(iv)}] Only agent \( i \)'s bundle changes within \( \ragents \), and her value for the new bundle increases. Thus, this property holds trivially.
		
		\item[\textsf{(v)}] The new bundle of agent \( i \) is a subset of \( \relevant{\pool}{i,j} \).
	\end{itemize}
\end{proof}

%

\subsection*{Update Rule 2}\label{Rule:2}

This rule applies when the corresponding vertex in $G_{0, \ralloc}$ for an agent $i \in \agents$ has an out-degree of 0. Agent $i$ must select the more valuable bundle between $\ralloc_i$ and $\relevant{\pool}{i}$ and return the other to the pool. Next, we finalize agent $i$'s bundle, and move her to $\sagents$.

\vspace{0.5cm}
\begin{algorithm}[H]
	\caption{Update Rule 2}
	\label{alg:rule2}
	\textbf{Input:} \(\ralloc,\ragents,\salloc,\sagents,\pool\)\;
	\If{\(\nexists\) agent in \(\ragents\) with out-degree 0 in \(G_{0,\ralloc}\)}{
		\Return \textit{(Not applicable)}\;
	}
	Let \(i\) be an agent in \(\ragents\) with out-degree 0 in \(G_{0,\ralloc}\)\;
	Let \(A \leftarrow \ralloc_i\)\;
	Let \(B \leftarrow \relevant{\pool}{i}\)\;
	\If{\(\valu_i(A) < \valu_i(B)\)}{
		\(\salloc_i \leftarrow B\)\;
		\(\pool \leftarrow (\pool \cup A)\setminus B\)\;
	}{
		\(\salloc_i \leftarrow A\)\;
	}
	Move agent \(i\) from \(\ragents\) to \(\sagents\)\;
	Remove bundle of \(i\) from \(\ralloc\)\;
\end{algorithm}

\begin{lemma}\label{lem:rule2}
	Rule 2 preserves properties $\textsf{(i)}$ to $\textsf{(v)}$.
\end{lemma}
\begin{proof}
	Suppose agent $i$ is the agent whose Rule 2 is applied on. We prove each property separately.
	
		\begin{itemize}
		\item[\textsf{(i)}] For each agent in $\sagents$, Property \textsf{(ii)} ensures that they did not $\sqrt{2}$-envy the combined set of goods in $\ralloc$ and $\pool$ before the update. Since the update only exchanges goods between agent $i$'s bundle and the pool, no agent in $\sagents$ $\sqrt{2}$-strongly envies agent $i$ afterward.
		Moreover, by Property \textsf{(iv)}, agent $i$ did not strongly envy any agent in $\sagents$ prior to the update, and her value has increased during the update. Thus, the resulting allocation $\salloc$ satisfies $(\nicefrac{1}{\sqrt{2}})$-$\efx$.
		
		\item[\textsf{(ii)}] We only need to verify this property for agent $i$, who is the sole agent added to $\sagents$ during the update. Since her out-degree in ${G_{0,\ralloc}}$ is zero, she assigns zero value to the bundles of all other agents in $\ragents$. Additionally, as she chooses the more valuable bundle between $\ralloc_i$ and $\relevant{\pool}{i}$, she will not envy the pool after the update. Combining these facts, agent $i$ does not envy the combined set of goods in $\ralloc$ and $\pool$.
		
		\item[\textsf{(iii)}] Agent~$i$ is no longer in $\ragents$, and the bundles of the remaining agents in $\ragents$ have not changed. Therefore, this property still holds.
		
		\item[\textsf{(iv)}] Since Rule~1 was not applicable when this rule was applied, no remaining agent envies $\relevant{\pool}{i}$. Moreover, by Property~\textsf{(iii)}, the allocation $\ralloc$ was $\efx$ before the update, so no agent in $\ragents$ strongly envied $\ralloc_i$. As the final bundle of agent $i$ is chosen from either $\ralloc_i$ or $\relevant{\pool}{i}$, and both were not strongly envied, it follows that no remaining agent strongly envies $i$'s final bundle.

		\item[\textsf{(v)}] Agent~$i$ is no longer in $\ragents$, and the bundles of the remaining agents in $\ragents$ have not changed. Therefore, this property still holds.
	\end{itemize}
	
\end{proof}

\begin{lemma}\label{lem:cycles}
	If the first two rules are not applicable, then $G_{0, \ralloc}$ consists of disjoint cycles. 
\end{lemma}
\begin{proof}
	By Property \textsf{(v)}, the in-degree of each vertex in $G_{0, \ralloc}$ is at most $1$. Since Rule 2 is not applicable, the out-degree of each vertex is at least $1$. Therefore $G_{0, \ralloc}$ is union of disjoint cycles. 
\end{proof}

\subsection*{Update Rule 3}\label{Rule:3}

By \Cref{lem:cycles}, \(G_{0,\ralloc}\) consists of several disjoint cycles. Rule 3 applies when there is a 2-cycle \(i \rightarrow j \rightarrow i\). We distinguish two cases: either \(i\) and \(j\) are the last two remaining agents, or there is at least one other agent.
For the first case, we invoke the following result by \citet{mahara2022twovaluations}:

\begin{theorem}[\citet{mahara2022twovaluations} --- Theorem 6 restated]\label{thm:two}
	Suppose \(X\) is a partial $\efx$ allocation and the agents have at most two distinct valuation functions.  Then one can extend \(X\) to an \(\efx\) allocation on all goods in \(\pool\) without decreasing any agent's utility.
\end{theorem}

Since here only agents $i$ and $j$ remain and they have at most two distinct valuation functions, this theorem guarantees we can allocate every good in $\pool$ while preserving $\efx$ between $i$ and $j$. \footnote{\cite{mahara2022twovaluations} assumes that each valuation is shared by at least two agents. For two agents, this assumption can be met by adding a dummy agent for each valuation and allocating them a dummy good with infinite value.}
Hence we can reallocate $\ralloc_i\cup\ralloc_j\cup\pool$ to $i$ and $j$ so that neither agent's utility decreases, finalize both bundles, and terminate the algorithm.

In the second case (at least three agents remain), we proceed as follows. If $i$ envies $j$ and vice versa, swap their bundles. Therefore, assume without loss of generality, $j$ no longer envies $i$. Next, partition the goods relevant to both $i$ and $j$ into three bundles 
$
\ralloc_i \cup (\relevant{\pool}{i}\setminus \relevant{\pool}{i,j}),
\ralloc_j,$ and 
$\relevant{\pool}{j}.
$
Return one of these three bundles to the pool and allocate the other two to $i$ and $j$ via the following choice protocol:
\begin{enumerate}
	\item Ask $j$ to choose between $\ralloc_j$ and $\relevant{\pool}{j}$.
	\item If $j$ picks $\ralloc_j$, give $i$ the bundle $\ralloc_i \cup (\relevant{\pool}{i}\setminus \relevant{\pool}{i,j})$.
	\item Otherwise, allocate $\relevant{\pool}{j}$ to $j$ and then ask $i$ to choose between $\ralloc_i \cup (\relevant{\pool}{i}\setminus \relevant{\pool}{i,j})$ and $\ralloc_j$.
\end{enumerate}
Finally, mark both $i$ and $j$ as satisfied, move them to $\sagents$, and remove their bundles from $\ralloc$.

\begin{algorithm}[h]
	\caption{Rule 3}
	\label{alg:rule3}
	\textbf{Input:} \(\ralloc,\ragents,\salloc,\sagents,\pool\)\;
	\If{\(\nexists\) cycle of length 2 in \(G_{0,\ralloc}\)}{
		\Return \textit{(Not Applicable)}\;
	}
	Let \(i \rightarrow j \rightarrow i\) be a cycle of length 2\;
	\If{\(|\ragents| = 2\)}{
		Allocate \(\ralloc_i \cup \ralloc_j \cup \pool\) between \(i,j\) \tcp*{using \Cref{thm:two} from \cite{mahara2022twovaluations}}
		Move \(i,j\) to \(\sagents\)\;
		\Return\;
	}
	\If{\(i\) envies \(\ralloc_j\) \textbf{and} \(j\) envies \(\ralloc_i\)}{
		Swap bundles: \(\ralloc_i \leftrightarrow \ralloc_j\)\;
	}
	\tcp{W.l.o.g., now \(j\) does not envy \(i\)}
	Let 
	\(A \leftarrow \ralloc_i \cup \bigl(\relevant{\pool}{i}\setminus \relevant{\pool}{i,j}\bigr)\),  
	\(B \leftarrow \ralloc_j\),  
	\(C \leftarrow \relevant{\pool}{j}\)\;
	\If{\(j\) prefers \(B\) over \(C\)}{
		Allocate \(B\) to \(j\)\;
		Allocate \(A\) to \(i\)\;
		\(\pool \leftarrow \pool \cup C\)\;
	}{
		Allocate \(C\) to \(j\): \(\ralloc_j \leftarrow C\)\;
		\(\pool \leftarrow \pool \cup B\)\;
		\If{\(i\) prefers \(A\) over \(B\)}{
			Allocate \(A\) to \(i\)\;
		}{
			Allocate \(B\) to \(i\)\;
			\(\pool \leftarrow \pool \cup A\)\;
		}
	}
	Move \(i\) and \(j\) from \(\ragents\) to \(\sagents\)\;
	Remove \(\ralloc_i\) and \(\ralloc_j\) from \(\ralloc\)\;
\end{algorithm}

\begin{lemma}\label{lem:rule3}
		Rule 3 preserves properties $\textsf{(i)}$ to $\textsf{(v)}$.
\end{lemma}
\begin{proof}
	We verify each property one by one.
	
	\begin{itemize}
		\item[\textsf{(i)}]
		Agents already in \(\sagents\) do not \(\sqrt{2}\)-strongly envy each other by Property \textsf{(i)}. Since Rule 3 does not decrease the utility of any finalized agent, they also do not \(\sqrt{2}\)-strongly envy \(i\) or \(j\), by Property \textsf{(iv)}. So it remains to check envy between \(i\) and \(j\).
		
		If \(i\) and \(j\) are the only agents left, their allocation is $\efx$ by \cref{thm:two}, and neither strongly envies the other. Otherwise, in the choice protocol, agent \(j\) receives either \(\ralloc_j\) or \(\relevant{\pool}{j}\). In the first case, \(i\) does not strongly envy \(\ralloc_j\) by Property \textsf{(iii)}. In the second, \(i\) does not envy \(\relevant{\pool}{j}\), since Rule 1 was not applicable. Then \(i\) receives either \(\ralloc_i \cup (\relevant{\pool}{i} \setminus \relevant{\pool}{i,j})\) or \(\ralloc_j\). In the first case, \(j\) did not envy \(\ralloc_i\) by assumption, and in the second case, it is \(j\)'s original bundle. Thus, after the update, neither agent strongly envies the other.
		
		\item[\textsf{(ii)}]
		Only agents \(i\) and \(j\) are finalized under this rule. If they are the only two remaining agents, then \(\ragents\) becomes empty, so the property holds trivially. Otherwise, the three bundles
		$
		\ralloc_i \cup (\relevant{\pool}{i} \setminus \relevant{\pool}{i,j}),
		\ralloc_j,$ and
		$\relevant{\pool}{j}
		$
		cover all goods relevant to \(i\) or \(j\). Two are assigned to \(i\) and \(j\), and the third is returned to the pool. Each agent receives a bundle they value at least as much as the one returned. So Property \textsf{(ii)} is satisfied.
		
		\item[\textsf{(iii)}]
		This property held before the update, and Rule 3 finalizes only \(i\) and \(j\), leaving other agents bundles unchanged. Hence, the property continues to hold.
		
		\item[\textsf{(iv)}]
		If only \(i\) and \(j\) remain, then \(\ragents = \emptyset\), and the property holds. Otherwise, consider any agent \(k \in \ragents\). We must check that \(k\) does not strongly envy any of:
		$
		\ralloc_j,
		\relevant{\pool}{j},$ and
		$\ralloc_i \cup (\relevant{\pool}{i} \setminus \relevant{\pool}{i,j}).
		$
		No agent envies \(\ralloc_j\) by Property \textsf{(iii)}. No agent envies \(\relevant{\pool}{j}\) since Rule 1 was not applicable. Finally, \(\ralloc_i\) and \(\relevant{\pool}{i}\) are relevant only to \(i\) and \(j\), and since Rule 1 was not applicable, no other agent envies \(\relevant{\pool}{i}\). Thus, \(k\) does not strongly envy any new bundle, and the property holds.
		
		\item[\textsf{(v)}]
		This property also held before the update and remains valid since Rule 3 affects only \(i\) and \(j\), and leaves the rest of \(\ragents\) unchanged.
	\end{itemize}
\end{proof}

\color{black}

\subsection*{Update Rule 4}\label{Rule:4}
If the first three rules are not applicable, we know that $G_{0,\ralloc}$ consists of disjoint cycles. We classify these cycles as either homogeneous or heterogeneous, as defined below:

\begin{definition}
	\label{homhet}
	A cycle $C \in G_{0,\ralloc}$ is homogeneous if either
		every edge $i \rightarrow j$ in $C$ is heavy, or
every edge $i \rightarrow j$ in $C$ is light.
	Otherwise, the cycle is heterogeneous.
\end{definition}
This rule applies when a homogeneous cycle $C$ exists in $G_{0,\ralloc}$. The allocation is updated in two steps:

\begin{enumerate}
	\item If all edges in $C$ are heavy, we rotate the bundles within the cycle: for each edge $i \rightarrow j$ in $C$, allocate $\ralloc_j$ to agent $i$. 
\item Next, we finalize the bundles of agents in \(C\) by allocating relevant goods from \(\pool\). For each good \(\good\) relevant to some agent in \(C\), we allocate it to an arbitrary relevant agent in $C$, unless \(\good\) is relevant to both \(i\) and \(j\) and there is an edge \(i \to j\) in the cycle. In that case, we allocate \(\good\) to agent \(i\).
\end{enumerate}
Algorithm \ref{alg:rule4} shows a pseudocode for our method in Rule 4.

\begin{algorithm}[t]
	\caption{Update Rule 4}
	\label{alg:rule4}
	\textbf{Input:} \(\ralloc,\ragents,\salloc,\sagents,\pool\)\;
	\If{\(\nexists\) homogeneous cycle in \(G_{0,\ralloc}\)}{
		\Return \textit{(Not Applicable)}; \tcp{No homogeneous cycle}
	}
	Let \(C \leftarrow\) a homogeneous cycle in \(G_{0,\ralloc}\)\;
	\If{\(C\) is heavy}{
		\For{each edge \(i \rightarrow j\) in \(C\)}{
			\(\ralloc_i \leftarrow \ralloc_j\); \tcp{Rotate bundles}
			\(C \leftarrow \mathrm{reverse}(C)\); \tcp{Cycle reversed}
		}
	}
	\For{each \(i \rightarrow j \in C\)}{
		\(B_i \leftarrow \relevant{\pool}{i,j}\)\;
	}
	\(B \leftarrow \bigcup_{i \in C} B_i\)\;
	\(T \leftarrow \{\,g \in \pool \setminus B \mid g \in \relevant{\pool}{i}\text{ for some }i\in C\}\)\;
	\For{each \(g \in T\)}{
		Let \(i \in C\) such that \(g \in \relevant{\pool}{i}\)\;
		\(B_i \leftarrow B_i \cup \{g\}\)\;
	}
	Move all agents in \(C\) from \(\ragents\) to \(\sagents\)\;
	\For{each \(i \in C\)}{
		\(\salloc_i \leftarrow \ralloc_i \cup B_i\)\;
	}
	Remove bundles of \(C\) from \(\ralloc\); \tcp{Bundles now in \(\salloc\)}
	Update \(\pool\)\;
\end{algorithm}

\begin{lemma}\label{lem:rule4}
	Rule 4 satisfies properties $\textsf{(i)}$ to $\textsf{(v)}$.
\end{lemma}
\begin{proof}
	We verify that each property is preserved:
	
	\begin{itemize}
		\item[\textsf{(i)}]
		Previously finalized agents (in $\sagents$) do not $\sqrt{2}$-strongly envy one another by Property \textsf{(i)}. Also, by Property \textsf{(iv)}, agents in the cycle $C$ do not strongly envy any finalized agent. Since their utilities decrease by at most a factor of $\sqrt{2}$ after the update, this remains true post-update.
		
		It remains to show that agents in $C$ do not $\sqrt{2}$-strongly envy each other after the update. Let $i$ be an agent in $C$, and consider the incoming edge $j \rightarrow i$ in the cycle after step (i) of Rule 4. By Property \textsf{(ii)}, $i$'s bundle is relevant only to $i$ and $j$.
		
		We claim that $j$ does not $\sqrt{2}$-envy $i$ after step (ii). First, note that no good from $\relevant{\pool}{j}$ is given to $i$ in this step. If all edges in $C$ are light, then $j$ does not envy $i$ at all. If all edges are heavy, then $i$ receives $j$'s previous bundle. Since $j \rightarrow i$ is a heavy edge, this guarantees that $j$ does not $\sqrt{2}$-envy $i$ after the rotation.
		
		For any other agent $k$ in $C$, Rule 1 not being applicable ensures that $k$ does not envy $\relevant{\pool}{i}$. Thus, agents in $C$ remain $\sqrt{2}$-envy-free among themselves.
		
		Finally, Property \textsf{(ii)} ensures that finalized agents do not $\sqrt{2}$-envy agents in $C$.
		
		\item[\textsf{(ii)}]
		Step (ii) allocates all goods relevant to $C$ to agents within $C$. As a result, no agent in $C$ envies the pool or any other agent's bundle--since nothing relevant is left unallocated.
		
		\item[\textsf{(iii)}]
		All agents in $C$ are removed from $\ragents$, and the bundles of the remaining agents are unchanged. Hence, this property continues to hold.
		
		\item[\textsf{(iv)}]
		It suffices to check that no agent in $\ragents$ strongly envies any agent in $C$. Before step (ii), bundles held by agents in $C$ are irrelevant to all agents outside $C$. Step (ii) only reallocates goods relevant to $C$, and since Rule 1 was not applicable beforehand, no agent in $\ragents$ envies any updated bundle in $C$. Therefore, this property is preserved.
		
		\item[\textsf{(v)}]
		As in \textsf{(iii)}, agents in $C$ are removed from $\ragents$, and all other bundles remain unchanged. Thus, this property continues to hold.
	\end{itemize}
	
	We conclude that Rule~4 preserves Properties \textsf{(i)} to \textsf{(v)}.
\end{proof}

\subsection*{Update Rule 5}\label{Rule:5}
The last case we handle with Rule 5 occurs when $G_{0,\ralloc}$ consists solely of heterogeneous cycles. For every cycle $C$, since $C$ is heterogeneous, there exist three consecutive agents $i \rightarrow j \rightarrow k$ in $G_{0, \ralloc}$, where $i \rightarrow j$ is heavy but $j \rightarrow k$ is light. We perform the update based on three agents as follows:

If $\valu_{j}(\pool) \le  \valu_{j}(\ralloc_{j})/\sqrt{2}$, we finalize the bundle of agent $j$ without any further changes. Otherwise, we allocate $\relevant{\pool}{j}$ to agent $j$ and $\ralloc_{j} \cup \relevant{\pool}{i} \setminus \relevant{\pool}{i, j}$ to agent $i$, then return $\ralloc_i$ to the pool and finalize the bundle of agents $i$ and $j$. 

\begin{algorithm}[H]
	\caption{Rule 5}
	\label{alg:rule5}
	\textbf{Input:} \(\ralloc,\ragents,\salloc,\sagents,\pool\)\;
	\If{\(\nexists\) heterogeneous cycle in \(G_{0,\ralloc}\)}{
		\Return \textit{(Not Applicable)}; \tcp{No heterogeneous cycle}
	}
	Let \(C \leftarrow\) a heterogeneous cycle in \(G_{0,\ralloc}\)\;
	Find \(i \rightarrow j \rightarrow k\) in \(C\) where \(i\rightarrow j\) is heavy and \(j\rightarrow k\) is light\;
	\If{\(\valu_j(\relevant{\pool}{j}) \le \valu_j(\ralloc_j)/\sqrt{2}\)}{
		Move \(j\) from \(\ragents\) to \(\sagents\); \tcp{Finalize with current bundle}
		Remove \(\ralloc_j\) from \(\ralloc\)\;
	}{
		\(\ralloc_j \leftarrow \relevant{\pool}{j}\); \tcp{Give relevant goods to \(j\)}
		\(\ralloc_i \leftarrow \ralloc_j \cup (\relevant{\pool}{i}\setminus \relevant{\pool}{i,j})\); \tcp{Adjust \(i\)'s bundle}
		Update \(\pool\)\;
		Move \(i,j\) from \(\ragents\) to \(\sagents\)\;
		Remove \(\ralloc_i, \ralloc_j\) from \(\ralloc\)\;
	}
\end{algorithm}

\begin{lemma}\label{lem:rule5}
	Rule 5 satisfies properties $\textsf{(i)}$ to $\textsf{(v)}$.
\end{lemma}
\begin{proof}
	We verify that each property is preserved:
	
	\begin{itemize}
		\item[\textsf{(i)}]
		We distinguish two cases based on agent \(j\)'s utility from the pool:
		
		\textbf{Case 1:} \(\valu_{j}(\relevant{\pool}{j}) \le \valu_{j}(\ralloc_{j}) / \sqrt{2}\).  
		In this case, only agent \(j\) is finalized with her current bundle.  
		By Property \textsf{(iii)}, she does not strongly envy any finalized agent.  
		By Property \textsf{(ii)}, no finalized agent \(\sqrt{2}\)-envies \(j\)'s bundle.
				
		\textbf{Case 2:} \(\valu_{j}(\relevant{\pool}{j}) > \valu_{j}(\ralloc_{j}) / \sqrt{2}\).  
		Here, both \(i\) and \(j\) are finalized. Since the edge \(i \rightarrow j\) is heavy, both agents' utilities drop by at most a factor of \(\sqrt{2}\).  
		Thus, neither \(i\) nor \(j\) \(\sqrt{2}\)-strongly envies any finalized agent.  
		Property \textsf{(ii)} ensures that no finalized agent \(\sqrt{2}\)-envies \(i\) or \(j\).  
		Since Rule 1 does not apply to \(\ralloc\), agent \(i\) does not \(\sqrt{2}\)-envy \(\relevant{\pool}{j}\).  
		And since \(j\)'s bundle is reassigned to \(i\), \(j\) does not \(\sqrt{2}\)-envy \(i\) after the update.
		
		\item[\textsf{(ii)}]
		We analyze the relevant bundles for agents \(i\) and \(j\):  
		\begin{itemize}
			\item For \(j\): \(\ralloc_j\), \(\ralloc_k\), and \(\relevant{\pool}{j}\).  
			\item For \(i\): \(\ralloc_i\), \(\ralloc_j\), and \(\relevant{\pool}{i}\).  
		\end{itemize}
		Again, we consider two cases:
		
		\textbf{Case 1:} \(\valu_{j}(\relevant{\pool}{j}) \le \valu_{j}(\ralloc_{j}) / \sqrt{2}\).  
		Only agent \(j\) is finalized. Since the edge \(j \rightarrow k\) is light:
		\[
		\valu_j(\relevant{\pool}{j}) + \valu_j(\ralloc_k) \le \invsq\, \valu_j(\ralloc_j) + \invsq\, \valu_j(\ralloc_j) = \sqrt{2}\, \valu_j(\ralloc_j).
		\]
		Hence, agent \(j\) does not \(\sqrt{2}\)-envy the union of remaining bundles.
		
		\textbf{Case 2:} \(\valu_{j}(\relevant{\pool}{j}) > \valu_{j}(\ralloc_{j}) / \sqrt{2}\).  
		Agent \(j\) receives \(\relevant{\pool}{j}\), and the only remaining relevant bundle is \(\ralloc_k\).  
		After the update, her utility drops by at most \(\sqrt{2}\), and since \(j \rightarrow k\) is light, she does not envy \(\ralloc_k\).  
		Agent \(i\) only finds \(\ralloc_i\) relevant among remaining bundles. Since the edge \(i \rightarrow j\) is heavy, she does not \(\sqrt{2}\)-envy the remaining bundles.
		
		\item[\textsf{(iii)}]
		Since the bundles of the remaining agents in \(\ragents\) are unchanged, this property is preserved.
		
\item[\textsf{(iv)}]
We show that no remaining agent strongly envies any newly finalized agent. If only \(j\) is finalized: her bundle remains unchanged, and since \(\ralloc\) satisfies Property \textsf{(iii)}, no remaining agent strongly envies \(j\). Furthermore, If both \(i\) and \(j\) are finalized:
Agent \(j\) receives \(\relevant{\pool}{j}\), and since Rule 1 does not apply, no remaining agent envies her.
Also, agent \(i\) receives \(\ralloc_j\), which is a subset of \(\relevant{\goods}{i,j}\), and thus irrelevant to the remaining agents.
		
		\item[\textsf{(v)}]
		Since the bundles of remaining agents are unchanged, this property is preserved.
	\end{itemize}
\end{proof}

\subsection*{Final Step}\label{UF}

Recall that in the first update rule, the social welfare of the agents strictly increases, and in each subsequent rule, the bundle of at least one agent is finalized. Therefore, after a finite number of updates, we reach a state where none of the rules are applicable, meaning all agents' bundles are finalized. However, when all bundles have been finalized, some goods may still remain unallocated in the pool. In the final step, these remaining goods are allocated to the last agent whose bundle was finalized.

\begin{theorem}\label{thm:efx}
	Suppose that the valuations are $(2, \infty)$-bounded, and let $\alloc^*$ be the allocation returned by the Algorithm \ref{alg1}. Then, $\alloc^*$ is a $(\nicefrac{1}{\sqrt{2}})$-$\efx$ allocation.
\end{theorem}
\begin{proof}
	Let $i$ be the last agent whose bundle was finalized. By Property $\textsf{(i)}$, the allocation was $(\nicefrac{1}{\sqrt{2}})$-$\efx$ for all agents prior to adding the remaining goods. Thus, the only possible $\sqrt{2}$-strongly envies in $\alloc^*$ are toward agent $i$. We consider cases based on the updating rule at which the algorithm terminates:
	\begin{itemize}
		\item The algorithm does not terminate with Rule 1 or 5. Rule 1 does not finalize any agent, and Rule 5 does not finalize all agents in the cycle. If the algorithm terminates with Rule 3, it is explicitly shown that the remaining goods can be divided between two agents such that the resulting allocation is $(\nicefrac{1}{\sqrt{2}})$-$\efx$.
		
		\item Suppose that the algorithm terminates at Rule 2. At that point, agent $i$ is the only remaining agent. Since the allocation satisfies Property $\textsf{(ii)}$, the finalized agents do not $\sqrt{2}$ envy toward the union of the pool and agent $i$'s bundle. Therefore, there is no $\sqrt{2}$-strongly envy toward agent $i$ in the final allocation.
		
		\item 
		
		Consider the case where the algorithm terminates at Rule 4. In this step, the remaining agents form a homogeneous cycle, and the rule allocates all remaining goods that are relevant to at least one of them so that no agent $\sqrt{2}$-envies another. As a result, the leftover goods are not relevant to these agents, and they do not $\sqrt{2}$-envy agent $i$. Moreover, the finalized agents also do not $\sqrt{2}$-envy agent $i$, since the allocation satisfies Property $\textsf{(ii)}$, and the finalized agents do not $\sqrt{2}$-envy the union of all remaining goods. Thus, no one $\sqrt{2}$-envies agent $i$.
	\end{itemize}
	Therefore, the agents do not $\sqrt{2}$-strongly envy each other and the allocation is $(\nicefrac{1}{\sqrt{2}})$-$\efx$.
\end{proof}

	\newpage
	\bibliographystyle{apalike}
	\bibliography{draft}
	\appendix
\end{document}